\documentclass[useAMS,usenatbib,usegraphicx,fleqn]{mn2e}

\usepackage[T1]{fontenc}
\usepackage{pslatex}
\usepackage{color}
\usepackage{amsmath}
\usepackage{url}
\usepackage{amssymb}
\def\gtrsim{\mathrel{\hbox{\rlap{\hbox{\lower3pt\hbox{$\sim$}}}\hbox{\raise2pt\hbox{$>$}}}}}

\definecolor{orange}{rgb}{1,0.3,0}
\definecolor{purple}{rgb}{1,0,1}

\newcommand{\nodata}{...}

\title[SALT observations of ACT SZ clusters]{SALT spectroscopic observations of galaxy clusters
detected by ACT and a Type II quasar hosted by a brightest cluster galaxy}
\author[Kirk et al.]
{\parbox{\textwidth}{\raggedright Brian~Kirk,$^{1,2}$\thanks{E-mail: bmarshallk@gmail.com}
Matt~Hilton,$^{1,3}$\thanks{E-mail: hiltonm@ukzn.ac.za}
Catherine~Cress,$^{2,4}$
Steven~M.~Crawford,$^{5}$
John~P.~Hughes,$^{6}$
Nicholas~Battaglia,$^{7}$
J.~Richard~Bond,$^{8}$
Claire~Burke,$^{1}$
Megan~B.~Gralla,$^{9}$
Amir~Hajian,$^{8}$
Matthew~Hasselfield,$^{10}$
Adam~D.~Hincks,$^{11}$
Leopoldo~Infante,$^{12}$
Arthur~Kosowsky,$^{13}$
Tobias~A.~Marriage,$^{9}$
Felipe~Menanteau,$^{14}$
Kavilan~Moodley,$^{1}$
Michael~D.~Niemack,$^{15}$
Jonathan~L.~Sievers,$^{16}$
Crist\'obal~Sif\'on,$^{17}$
Susan~Wilson,$^{1}$
Edward~J.~Wollack,$^{18}$
and
Caroline~Zunckel$^{16}$
}\vspace{0.4cm}\\
\parbox{\textwidth}{\raggedright 
$^{1}$~Astrophysics \& Cosmology Research Unit, School of Mathematics, Statistics \& Computer Science, University of KwaZulu-Natal, Durban 4041, SA\\
$^{2}$~Centre for High Performance Computing, CSIR Campus, 15 Lower Hope St. Rosebank, Cape Town, SA\\
$^{3}$~Centre for Astronomy \& Particle Theory, School of Physics and Astronomy, University of Nottingham, NG7 2RD, UK\\
$^{4}$~Department of Physics, University of Western Cape, Bellville 7530, Cape Town, SA\\
$^{5}$~South African Astronomical Observatory, P.O. Box 9, Observatory 7935, Cape Town, SA\\
$^{6}$~Department of Physics and Astronomy, Rutgers, The State University of New Jersey, 136 Frelinghuysen Road, Piscataway, NJ USA 08854-8019\\
$^{7}$~McWilliams Center for Cosmology, Carnegie Mellon University, Department of Physics, 5000 Forbes Ave., Pittsburgh PA, USA, 15213\\
$^{8}$~Canadian Institute for Theoretical Astrophysics, 60 St George, Toronto ON, Canada, M5S 3H8\\
$^{9}$~Department of Physics and Astronomy, Johns Hopkins University, 3400 N. Charles St., Baltimore, MD 21218\\
$^{10}$~Department of Astrophysical Sciences, Peyton Hall, Princeton University, Princeton, NJ 08544, USA\\
$^{11}$~Department of Physics and Astronomy, University of British Columbia, 6224 Agricultural Road, Vancouver, BC V6T 1Z1, Canada\\
$^{12}$~Departamento de Astronom\'ia y Astrof\'isica, Pontific\'ia Universidad Cat\'olica, Casilla 306, Santiago 22, Chile\\
$^{13}$~Department of Physics and Astronomy, University of Pittsburgh, Pittsburgh, PA 15260 USA\\
$^{14}$~National Center for Supercomputing Applications, University of Illinois at Urbana-Champaign,  1205 W. Clark St, Urbana, IL 61801\\
$^{15}$~Department of Physics, Cornell University, Ithaca, NY 14853, USA\\
$^{16}$~Astrophysics \& Cosmology Research Unit, School of Chemistry \& Physics, University of KwaZulu-Natal, Durban 4041, SA\\
$^{17}$~Leiden Observatory, Leiden University, PO Box 9513, NL-2300 RA Leiden, Netherlands\\
$^{18}$~NASA/Goddard Space Flight Center, Greenbelt, MD, 20771, USA\\
}
}

\begin{document}

\renewcommand{\bottomfraction}{0.95}

\date{Accepted for publication in MNRAS}

\pagerange{\pageref{firstpage}--\pageref{lastpage}} \pubyear{2015}

\maketitle

\label{firstpage}

\begin{abstract}

We present Southern African Large Telescope (SALT) follow-up observations of seven massive clusters detected 
by the Atacama Cosmology Telescope (ACT) on the celestial equator using the Sunyaev-Zel'dovich (SZ) effect. We conducted 
multi-object spectroscopic observations with the Robert Stobie Spectrograph in order to measure galaxy redshifts
in each cluster field, determine the cluster line-of-sight velocity dispersions, and infer the cluster dynamical masses. 
We find that
the clusters, which span the redshift
range $0.3 < z < 0.55$, range in mass from $(5 - 20) \times 10^{14}$\,M$_{\sun}$ ($M_{200c}$). 
Their masses, given their SZ signals, are similar to those of 
southern hemisphere ACT clusters previously observed using Gemini and the VLT. We note that the 
brightest cluster galaxy in one of the systems studied, ACT-CL~J0320.4+0032 at $z = 0.38$,
hosts a Type II quasar. Only a handful of such systems are currently known, and 
therefore ACT-CL~J0320.4+0032 may be a rare example of a very massive halo in which quasar-mode feedback is actively 
taking place.
\end{abstract}


\section{Introduction}
\label{s_intro}

Clusters of galaxies mark the highest density regions of the Universe at mega parsec scales. By charting the evolution of 
their number density as a function of mass and redshift, one is able to obtain constraints on cosmological
parameters, including the amount of dark matter and dark energy in the Universe 
\citep[e.g.,][]{Vikhlinin_2009, Mantz_2010, Sehgal_2011, Benson_2013, Hasselfield_2013, Planck_XX_2013}. However, cluster mass
-- which is the property predicted by $N$-body simulations of cold dark matter -- is 
not a directly measurable quantity, and must instead be inferred from the observable properties of the clusters. 
This has led to many studies that derive mass--observable scaling relations using a wide 
variety of observables including, optical richness \citep[e.g.,][]{Rozo_2009}; 
X-ray luminosity and temperature \citep[e.g.,][]{Vikhlinin_2006}; and Sunyaev-Zel'dovich effect signal 
\citep[e.g.,][]{Sifon_2013}.

The discovery of new clusters from large area surveys using the Sunyaev-Zel'dovich effect \citep[SZ;][]{SZ_1970}
began only in recent years \citep[e.g.,][]{Staniszewski_2009, Vanderlinde_2010, Marriage_2011, Reichardt_2013, Planck_XXIX_2013, Bleem_2014}. 
The SZ effect is the inverse Compton scattering of cosmic microwave background photons by hot ($> 10^7$\,K) 
gas trapped within the deep gravitational potential wells of massive galaxy clusters. It is almost redshift
independent, and in principle, it allows the discovery of all clusters in the Universe above a mass limit
set by the noise properties of the SZ survey \citep[see, e.g.,][]{Birkinshaw_1999, Carlstrom_2002}. In
addition, the SZ signal, usually denoted by the integrated Comptonisation ($Y$) parameter, has been shown
to correlate with cluster mass, with relatively small scatter 
\citep[e.g.,][]{Planck_XI_2013, Hoekstra_2012, Sifon_2013}. Despite this, mass--calibration is the main
contribution to the error budget of current cosmological studies using SZ-selected cluster samples
\citep[e.g.,][]{Sehgal_2011, Hasselfield_2013, Reichardt_2013, Planck_XX_2013, Bocquet_2014}, 
and so further work in this area is clearly needed.

In this paper, we present the results of a pilot follow-up study of SZ-selected clusters detected by the 
Atacama Cosmology Telescope \citep[ACT;][]{Swetz_2011} conducted using the Robert Stobie Spectrograph
\citep[RSS;][]{Burgh_2003} on the Southern African Large Telescope \citep[SALT;][]{Buckley_2006}. The goals
of this programme were to obtain spectroscopic redshifts and dynamical mass estimates through velocity
dispersions for ACT clusters, with the aim of increasing the sample of clusters for our
calibration of the $Y$-mass relation (\citealt{Sifon_2013}; see \citealt{Hasselfield_2013} 
for joint constraints on the dynamical mass scaling relation and cosmological parameters). 

The structure of this paper is as follows. We briefly describe the ACT cluster sample and the design, 
execution, and reduction of the SALT spectroscopic observations in Section~\ref{s_data}.
Section~\ref{s_results} presents the cluster redshifts and velocity dispersions. 
We compare the properties of the clusters studied here to previous observations of SZ clusters in
Section~\ref{s_discussion} and summarise our findings in Section~\ref{s_conclusions}.

We assume a cosmology with $\Omega_{\rm m}=0.3$, $\Omega_\Lambda=0.7$, and
$H_0=70$~km~s$^{-1}$~Mpc$^{-1}$ throughout. All magnitudes are on the AB system \citep{Oke_1974}, unless
otherwise stated.

\section{Observations and Analysis}
\label{s_data}

\subsection{Cluster sample}
\label{s_sample}
The clusters targeted for SALT observations were drawn from the SZ-selected sample constructed by the ACT 
team \citep[][]{Hasselfield_2013, Menanteau_2013}. ACT \citep{Swetz_2011} is a 6\,m telescope located in northern Chile that observes the sky in
three frequency bands (centred at 148, 218, and 277 GHz) simultaneously with arcminute resolution.

ACT surveyed two regions of the sky, searching a total area of 959 deg$^{2}$ for SZ galaxy clusters. 
During 2008, ACT observed a 455 deg$^{2}$ patch of the Southern sky,
centred on $\delta = -55$ deg, detecting a total of 23 massive clusters that were optically
confirmed using 4\,m class telescopes \citep{Menanteau_2010, Marriage_2011}. From 2009--2010, ACT 
observed a 504\,deg$^2$ region centred on the celestial equator, an area chosen due to its complete overlap with the deep 
($r \approx$ 23.5 mag) optical data from the 270\,deg$^2$ Stripe 82 region \citep[][]{Annis_2012}
of the Sloan Digital Sky Survey 
\citep[SDSS;][] {Abazajian_2009}.
\citet{Hasselfield_2013} describes the construction of the SZ cluster sample from the 148 GHz maps 
(see \citealt{Dunner_2013} for a detailed description of the reduction of the ACT data from 
timestreams to maps). Optical confirmation and redshifts for these clusters are reported in 
\citet{Menanteau_2013}, using data from SDSS and additional targeted optical and IR observations 
obtained at Apache Point Observatory. All 68 clusters in the sample have either photometric 
redshift estimates, or spectroscopic redshifts (largely derived from SDSS data). The sample spans
the redshift range $0.1 < z < 1.4$, with median $z = 0.5$.

In this pilot study with SALT, we targeted seven of the equatorial ACT clusters detected with reasonably
high signal-to-noise ($4.6 < S/N < 8.3$) and at moderate redshift ($z \approx 0.4$), in order to ensure that 
targeted galaxies would be bright enough for successful absorption line redshift measurements given the
capabilities of the RSS instrument at the time of the observations\footnotemark. Sif\'on et al. (in prep.)
will present observations of a further 21 ACT equatorial clusters observed with the Gemini telescopes.
\footnotetext{At the time of writing (September 2014), RSS is undergoing refurbishment that aims to 
increase its throughput considerably.}

\subsection{Spectroscopic observations}
\label{s_specObs}

\begin{table*}
\caption{Details of spectroscopic observations reported in this work. For all observations the pg0900 grating was used with the pc03850
order blocking filter. The CS and GA columns indicate the RSS camera station and grating angle used
respectively. Slitlets in all masks were 1.5$\arcsec$ wide and 10$\arcsec$ long. The position angle for all
observations was 180$\degr$, as all targets are on the celestial equator. The number of slits ($N_{\rm slits}$) does not
include alignment stars. Time allocations by SALT partner: RSA 50 per cent, Rutgers 50 per cent.}
\label{t_obsLog}
\begin{tabular}{|c|c|c|c|c|c|c|c|c|l|}
\hline

Program & Target & Mask & $N_{\rm slits}$ &  Frames & CS    & GA    & Airmass & Seeing & Date(s)\\
        &        &      &                 & (sec)   & (deg) & (deg) &         & (arsec)& (UT)\\            
\hline
2012-1-RSA\_UKSC\_RU-001 & J2058.8+0123 & 1 & 23 & $2 \times 975$ & 28.75 & 14.375 & 1.8 & 1.6 & 2012 Jul 16\\
2012-1-RSA\_UKSC\_RU-001 & J2058.8+0123 & 2 & 22 & $2 \times 975$ &  28.75 & 14.375 & 1.3 & 1.5 & 2012 Jul 24\\
2012-1-RSA\_UKSC\_RU-001 & J2058.8+0123 & 3 & 22 & $2 \times 975$ &  28.75 & 14.375 & 1.3 & 1.4 & 2012 Sep 06\\
2012-2-RSA\_UKSC\_RU-001 & J0320.4+0032 & 1 & 25 & $2 \times 975$ & 28.75 & 14.375 & 1.3& 1.3 & 2012 Nov 10\\
2012-2-RSA\_UKSC\_RU-001 & J0320.4+0032 & 2 & 26 & $4 \times 975$ & 28.75 & 14.375 & 1.3,1.3&0.9,1.4 & 2012 Nov 13, 15\\
2012-2-RSA\_UKSC\_RU-001 & J0320.4+0032 & 3 & 22 & $4 \times 975$ & 28.75 & 14.375 & 1.3 & 1.3 & 2012 Nov 16\\
2012-2-RSA\_UKSC\_RU-001 & J0219.9+0129 & 1 & 22 & $4 \times 975$ & 28.75 & 14.375 & 1.3,1.3& 1.4, 1.3 & 2012 Nov 15, 16\\
2013-1-RSA\_RU-001 & J0045.2-0152 & 1 & 26 & $4 \times 975$ & 32.50 & 16.250 & 1.2,1.2& 1.2,1.4 & 2013 Sep 01, 05\\
2013-1-RSA\_RU-001 & J0045.2-0152 & 2 & 25 & $2 \times 975$ & 32.50 & 16.250 & 1.2 & 1.5 & 2013 Sep 25\\
2013-2-RSA\_RU-002 & J0156.4-0123 & 1 & 25 & $2 \times 975$ & 31.00 & 15.50 & 1.2 & 1.6 & 2013 Nov 02\\
2013-2-RSA\_RU-002 & J0156.4-0123 & 3 & 21 & $2 \times 975$ & 31.00 & 15.50  & 1.3 & 1.3 & 2014 Jan 03\\
2013-2-RSA\_RU-002 & J0348.6-0028 & 1 & 25 & $2 \times 975$ & 28.75 & 14.375 & 1.2 & 1.4 & 2013 Nov 03\\
2013-2-RSA\_RU-002 & J0348.6-0028 & 2 & 23 & $2 \times 975$ & 28.75 & 14.375 & 2.0 & 1.9 & 2014 Jan 01\\
2013-2-RSA\_RU-002 & J0348.6-0028 & 3 & 23 & $4 \times 975$ & 28.75 & 14.375 & 1.3,1.2& 1.5, 0.8 & 2013 Nov 04, 08\\
2013-2-RSA\_RU-002 & J0348.6-0028 & 4 & 19 & $4 \times 975$ & 28.75 & 14.375 & 1.3,1.3& 1.4,1.5 & 2013 Dec 29, 30\\
2013-2-RSA\_RU-002 & J0342.7-0017 & 1 & 22 & $2 \times 975$ & 28.00 & 14.000 & 1.2 & 1.3 & 2014 Jan 03\\
\hline
\end{tabular}
\end{table*}

We conducted observations of the seven target ACT clusters with RSS in multi-object spectroscopy 
(MOS) mode, which uses custom designed slit masks. Given that SALT is located at Sutherland where the 
median seeing is $1.3\arcsec$ \citep{Catala_2013}, we chose to 
use slitlets with dimensions of 1.5$\arcsec$ width and 
10$\arcsec$ length. The latter was chosen to ensure reasonably accurate sky subtraction given these seeing
conditions. The RSS has an $8\arcmin$ diameter circular field of view, and with these slit dimensions we
found we were able to target 19-26 galaxies in each cluster field per slit mask. We selected 3-4 bright
($15 - 17.5$ magnitude in the $r$ band) stars per cluster field for alignment of the slit masks during 
acquisition.

The slit masks were designed using catalogues extracted from the 8th data release of the 
SDSS \citep[SDSS;][]{Aihara_2011}. We centred each slit mask on the Brightest
Cluster Galaxy (BCG) coordinates listed in \citet{Menanteau_2013} and estimated the colour of the 
red-sequence from visual inspection of the colour-magnitude diagrams. We used this information to define target
galaxy samples for each cluster, prioritising the selection of galaxies with magnitudes fainter than the BCG
and with colour bluer than our estimate of the red-edge of the red-sequence (note that these 
colour - magnitude cuts vary from cluster-to-cluster due to their slightly different redshifts). We then 
proceeded to assign slits to target 
galaxies in an automated fashion using an algorithm that prioritised objects closer to the cluster centre 
(in practice, this ensured that the number of objects whose spectra were centred horizontally on the detector
array was maximised). The final masks were made using the \textsc{PySlitMask} tool, part of the 
\textsc{PySALT}\footnotemark software package \citep{Crawford_2010}. We designed multiple masks for each 
target, although not all masks were observed.

\footnotetext{The PySALT user package is the primary reduction and analysis software tools for the SALT telescope (\url{http://pysalt.salt.ac.za/}).}

The RSS observations were conducted using the pg0900 Volume Phase Holographic (VPH) grating. We set the
RSS camera station and grating angle to centre the wavelength coverage at the expected wavelength of D4000 
for each cluster, since each cluster had either a spectroscopic or photometric redshift measurement 
\citep{Menanteau_2013}. The observing set up for $z \approx 0.3$
clusters (i.e., most clusters in this sample; camera station 28.75$\degr$, grating angle 14.375$\degr$) results
in dispersion 0.98\,\AA{} per binned pixel (2$\times$2 binning) with $4000-7000$\,\AA{} 
wavelength coverage. This results in a resolution of $\sim$ 4 \AA{}. There are two gaps in the spectral 
coverage due to physical gaps between three CCD chips that read out the dispersed spectra. 

The design of SALT limits observations of objects on the celestial equator to 
approximately 3200\,sec long intervals (referred to as observing blocks or tracks). In each observing block
the position of the tracker must be reset and the object re-acquired, the mask must be aligned, and flats and arcs
must be obtained. These operations incur significant overhead ($\approx$ 1200\,sec in total per block). We 
therefore obtained $2 \times 975$\,sec RSS exposures per observing block for our first observations in 
July-September 2012. For some subsequent observations, we obtained $4 \times 975$\,sec exposures by
observing each mask in two observing blocks. Note that SALT is a queue-scheduled telescope and observations
were obtained (sometimes of the same mask) on different nights throughout each observing semester. 
Table~\ref{t_obsLog} presents a summary of the observations.

\subsection{Spectroscopic data reduction}
\label{s_dataReduction}
A combination of \textsc{PySALT} and \textsc{IRAF}\footnotemark tasks were used to reduce the spectra. \textsc{PySALT} is a 
suite of \textsc{PyRAF} tools for the reduction and analysis of data obtained from the RSS instrument mounted on SALT 
(see \citealt{Crawford_2010}). 
\textsc{PySALT} tasks were used to prepare the image headers for the pipeline; apply CCD amplifier gain
and crosstalk corrections; subtract bias frames; perform cleaning of cosmic-rays; apply flat-field corrections; 
create mosaic images; and extract the data for each target based on the slit mask geometry.
IRAF tasks were then used to determine a wavelength dispersion function from a calibration lamp (Xenon or Argon); 
fit and transform the arc dispersion to the science frames; apply a background subtraction to each slitlet, the value of 
which is determined by a constant sampling area across the dispersion axis; combine images; and extract one dimensional spectra. For combined 
images, a maximum wavelength shift of 0.2\,\AA{} was measured between nights for observations of the same objects, well 
within the spectrograph resolution.

\footnotetext{\texttt{IRAF} is distributed by the National Optical Astronomy Observatories, which are operated
by the Association of Universities for Research in Astronomy, Inc., under cooperative agreement with the
National Science Foundation.}

\begin{figure*}
\includegraphics[width=17.5cm]{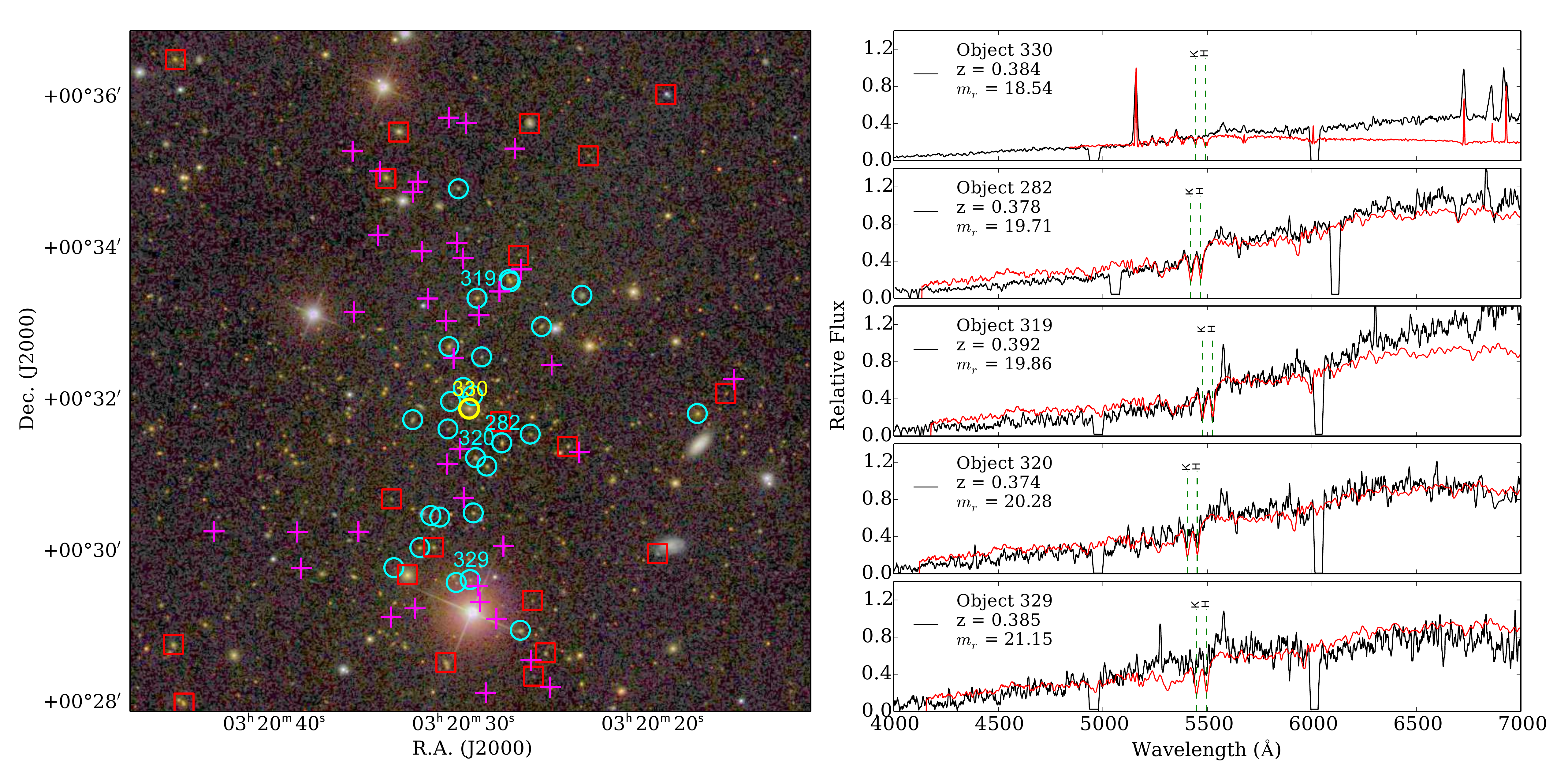}
\caption{The $z = 0.38$ cluster ACT-CL J0320.4+0032. The left hand panel shows a $9 \arcmin \times 9 \arcmin$
false colour SDSS optical image ($g$, $r$, $i$). Objects highlighted with cyan circles are spectroscopically confirmed
members (see Section~\ref{s_membership}; spectra for objects marked with ID numbers are shown in the
right hand panel);
red squares mark non-members with confirmed redshifts; and magenta crosses mark objects for which we failed to measure
a secure redshift ($Q < 3$; see Section~\ref{s_redshifts}). In the right hand 
panel, black lines correspond to SALT RSS spectra (smoothed with a 10 pixel boxcar), while red lines show the best match redshifted 
SDSS spectral template in each case. The displayed object spectra span a representative range in $r$-band
magnitudes, as indicated in the figure, and the spectrum for the brightest object is that of the BCG. In this
case, the BCG (object 330) is a Type II quasar (see Section~\ref{s_QSO}; it is highlighted with the yellow
circle in the image). Similar figures for the other clusters
can be found in the Appendix.}
\label{f_J0320ImageAndSpectra}
\end{figure*}

\begin{figure*}
\includegraphics[width=17.5cm]{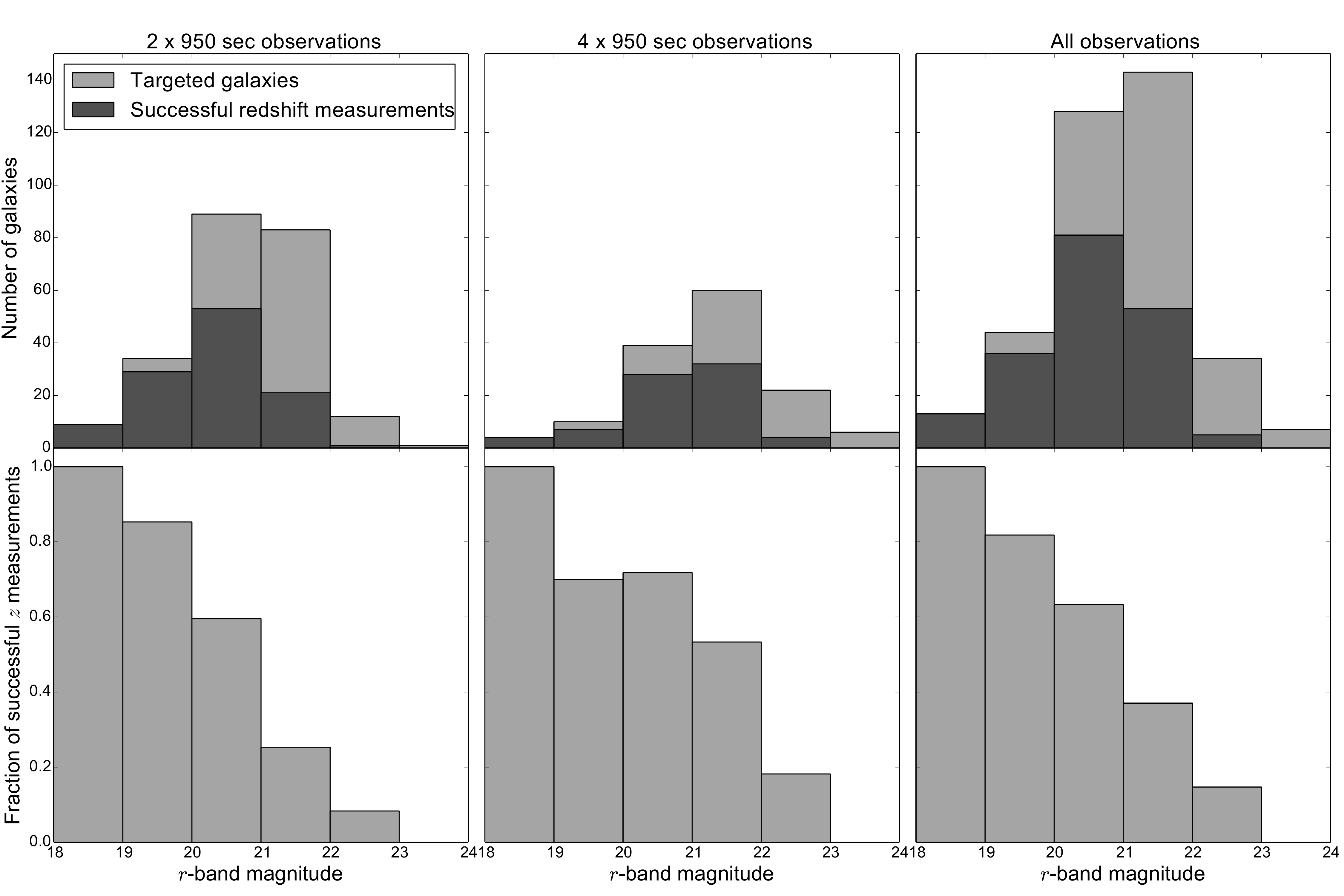}
\caption{Redshift success rate as a function of $r$-band magnitude, where a successful redshift measurement
is defined as having quality flag $Q \geq 3$. Results for $2 \times 975$\,sec of integration (one track), 
$4 \times 975$\,sec of integration (two tracks), and all observations are shown in the columns. 
Note that no attempt is made to control for the effect of 
variation in observing conditions.}
\label{f_successRecovery}
\end{figure*}


\subsection{Galaxy redshift measurements}
\label{s_redshifts}
Galaxy redshifts were measured by cross-correlating the spectra with SDSS galaxy spectral 
templates\footnotemark using the \textsc{RVSAO/XCSAO} package for \textsc{IRAF} 
\citep{KurtzMink_1998}. We ran the cross-correlation repeatedly with starting redshifts spanning 
$0.0 < z < 1$ in intervals of $\delta z = 0.0001$ for six different templates. We selected the redshift with
the highest correlation coefficient as the best measurement for the given template. This method provided 
six possible redshifts per galaxy spectrum. The final redshift measurement for each galaxy was selected 
from these six candidate redshifts after visual inspection of the 1d and 2d spectra by two or more of the 
co-authors. 
\footnotetext{\url{http://www.sdss.org/dr7/algorithms/spectemplates/index.html}}

We defined a quality rating system ($Q$) to describe the confidence level of each redshift 
measurement \citep[e.g.,][]{Wirth_2004}. Galaxies exhibiting multiple absorption and/or emission features
were given a $Q = 4$ rating; $Q = 3$ corresponds to galaxies exhibiting a single, strongly detected feature; 
galaxies showing the proper $z$ range but exhibiting no strong features were $Q = 2$; and 
galaxies with clearly spurious $z$ values (where the cross correlation failed due to poor signal-to-noise) 
and no strong features were rated $Q = 1$. Redshift measurements with $Q < 3$ were as a result of poor 
signal-to-noise spectra, slits blocked by the guide probe, or telescope malfunctions such as slit mask 
alignment failure resulting in manual alignment.

Spectra of members of each cluster, overplotted with best match spectral templates at the measured redshifts, 
are presented in Fig.~\ref{f_J0320ImageAndSpectra} (for ACT-CL~J0320.4+0032) and 
Figs.~A1--A6 in the Appendix for the other clusters in the sample. In these figures, the left hand panel shows a 
$9 \arcmin \times 9 \arcmin$ false colour SDSS optical image ($g$, $r$, $i$) of the cluster, highlighting
galaxies for which redshifts were measured. In the right hand panel, a selection of spectra spanning the 
magnitude range of the members are shown, and the spectrum for the brightest object is that of the BCG. 
In the case of ACT-CL~J0320.4+0032, we see relatively broad emission lines in the spectrum of the BCG
(Fig.~\ref{f_J0320ImageAndSpectra}). 
As discussed in Section \ref{s_discussion}, this galaxy is a Type II quasar host, and 
this may be a rare example of a massive cluster in which quasar-mode feedback is observed to be actively taking place.

Since all of our clusters are located within the SDSS footprint, we were able to verify the SALT redshift
measurements using a small number of objects in common with SDSS DR10 \citep{Ahn_2014}. From 9 
overlapping galaxies, we found that the median $\delta z = z_{\rm SALT} - z_{\rm SDSS} = -1.0 \times 10^{-5}$,
with standard deviation $\sigma = 3.9 \times 10^{-4}$ (we take the latter to indicate the level of 
uncertainty in the SALT redshift measurements). At $z = 0.3$, these translate into a 
median rest-frame velocity offset of $-2.3$\,km\,s$^{-1}$ with $\sigma = 90$\,km\,s$^{-1}$. Given the 
relatively low redshift of the clusters in this study, a search of SDSS DR10 also yielded some 
additional spectroscopic redshifts within most of the clusters that were not matched with galaxies 
targeted by SALT. These objects are included in the analysis presented in Section~\ref{s_results} below.

\subsection{Redshift success rate}
\label{s_SALT}
Since this project is one of the first to use the MOS mode of RSS to collect galaxy redshifts,
here we quantify our efficiency for the benefit of others planning to use this instrument for similar work.
Fig.~\ref{f_successRecovery} shows the redshift measurement success rate as a function of galaxy $r$-band 
magnitude, where we define a successful redshift measurement as one with $Q \geq 3$ (note that only 
galaxies with $Q \geq 3$ are included in the sample used to measure cluster velocity dispersions, 
as described in Section~\ref{s_results} below). Overall, we successfully measured redshifts for 191 out of 372 
galaxies targeted (51 per cent), spanning the $r$-band magnitude range 17.9--23.9; the top panels in 
Fig.~\ref{f_successRecovery} show the magnitude distributions of the target galaxies.

As described in Section~\ref{s_specObs} above, the design of SALT limits observations on the celestial
equator to tracks of 3200\,sec duration, and we observed some masks with one SALT track per target (obtaining 
$2 \times 975$\,sec of integration per mask), and others with two SALT tracks per target (obtaining
$4 \times 975$\,sec of integration per mask). The columns of Fig.~\ref{f_successRecovery} show how the 
redshift success rate changes depending on whether one or two tracks were used. We see that the 
two-track observations result in more than double the efficiency for measuring 
redshifts for galaxies with $21 < r$-band~mag~$< 23$. In two-track observations, we successfully
measured redshifts for 53 per cent of galaxies with $21 < r$-band~mag~$< 22$ and 18 per cent of galaxies
with $22 < r$-band~mag~$< 23$. The magnitude limit corresponding to redshift measurement efficiency of 
70 per cent is $r < 21$ for one track, compared to $r < 21.4$ for the two track observations. Note that none of the
above estimates take into account possible variation in observing conditions between the one versus 
two track observations, although the seeing was similar (see Table~\ref{t_obsLog}).


\section{Results}
\label{s_results}

\begin{table*}
\caption{Spectroscopic redshifts of galaxies in the direction of ACT-CL~J0320.4+0032 measured using SALT RSS:
$m_{r}$ is the SDSS $r$-band magnitude of the object; $z$ is the redshift; $Q$ is the redshift quality flag 
(see Section~\ref{s_redshifts}); Em. Lines? indicates objects which show emission lines in their spectra
(e.g., [\textsc{Oii}]\,$\lambda$\,3727); Member? indicates objects which are determined to be cluster members (see
Section~\ref{s_membership}); $r$ (Mpc) indicates the projected
distance from the BCG position as given in Menanteau~et~al.~(2013). Member galaxies in Mask 'S' have redshifts
from SDSS DR10 (Ahn~et~al.~2013). Similar tables for the other clusters are found in the Appendix.}
\label{t_members_J0320.4+0032}
\begin{tabular}{|c|c|c|c|c|c|c|c|c|c|}
\hline
ID & Mask & RA (J2000) & Dec. (J2000) & $m_{r}$ & $z$ & $Q$ & Em. Lines? & Member? & $r$ (Mpc)\\
\hline
330 & 1 & $03^{\rmn h} 20^{\rmn m} 29\fs788$ & $+00\degr 31\arcmin 53\farcs60$ & 18.54 & 0.3836 & 4 & $\checkmark$ & $\checkmark$ & 0.00\\
324 & 2 & $03^{\rmn h} 20^{\rmn m} 29\fs602$ & $+00\degr 32\arcmin 03\farcs99$ & 21.73 & 0.3884 & 4 & \nodata & $\checkmark$ & 0.05\\
356 & 3 & $03^{\rmn h} 20^{\rmn m} 30\fs772$ & $+00\degr 31\arcmin 59\farcs27$ & 21.92 & 0.3943 & 4 & \nodata & $\checkmark$ & 0.09\\
10 & S & $03^{\rmn h} 20^{\rmn m} 30\fs096$ & $+00\degr 32\arcmin 10\farcs41$ & 19.69 & 0.3883 & 4 & \nodata & $\checkmark$ & 0.09\\
360 & 1 & $03^{\rmn h} 20^{\rmn m} 30\fs907$ & $+00\degr 31\arcmin 37\farcs50$ & 21.75 & 0.3827 & 3 & \nodata & $\checkmark$ & 0.13\\
286 & 2 & $03^{\rmn h} 20^{\rmn m} 28\fs141$ & $+00\degr 31\arcmin 43\farcs24$ & 22.06 & 0.4725 & 3 & \nodata & \nodata & \nodata\\
282 & 2 & $03^{\rmn h} 20^{\rmn m} 28\fs066$ & $+00\degr 31\arcmin 26\farcs50$ & 19.71 & 0.3778 & 4 & \nodata & $\checkmark$ & 0.19\\
320 & 2 & $03^{\rmn h} 20^{\rmn m} 29\fs449$ & $+00\degr 31\arcmin 14\farcs78$ & 20.28 & 0.3737 & 4 & \nodata & $\checkmark$ & 0.20\\
311 & 2 & $03^{\rmn h} 20^{\rmn m} 29\fs130$ & $+00\degr 32\arcmin 34\farcs61$ & 21.02 & 0.3790 & 4 & $\checkmark$ & $\checkmark$ & 0.22\\
396 & 3 & $03^{\rmn h} 20^{\rmn m} 32\fs770$ & $+00\degr 31\arcmin 44\farcs87$ & 20.32 & 0.3842 & 4 & \nodata & $\checkmark$ & 0.24\\
303 & 3 & $03^{\rmn h} 20^{\rmn m} 28\fs853$ & $+00\degr 31\arcmin 08\farcs07$ & 20.42 & 0.3867 & 4 & \nodata & $\checkmark$ & 0.25\\
251 & 3 & $03^{\rmn h} 20^{\rmn m} 26\fs561$ & $+00\degr 31\arcmin 33\farcs46$ & 21.44 & 0.3841 & 4 & $\checkmark$ & $\checkmark$ & 0.27\\
6 & S & $03^{\rmn h} 20^{\rmn m} 30\fs857$ & $+00\degr 32\arcmin 42\farcs80$ & 19.46 & 0.3939 & 4 & \nodata & $\checkmark$ & 0.27\\
323 & 1 & $03^{\rmn h} 20^{\rmn m} 29\fs571$ & $+00\degr 30\arcmin 30\farcs98$ & 20.50 & 0.3830 & 4 & \nodata & $\checkmark$ & 0.43\\
238 & 3 & $03^{\rmn h} 20^{\rmn m} 25\fs962$ & $+00\degr 32\arcmin 58\farcs63$ & 20.62 & 0.3791 & 4 & \nodata & $\checkmark$ & 0.45\\
319 & 1 & $03^{\rmn h} 20^{\rmn m} 29\fs372$ & $+00\degr 33\arcmin 21\farcs18$ & 19.86 & 0.3922 & 4 & \nodata & $\checkmark$ & 0.46\\
368 & 2 & $03^{\rmn h} 20^{\rmn m} 31\fs339$ & $+00\degr 30\arcmin 27\farcs66$ & 21.22 & 0.3791 & 3 & $\checkmark$ & $\checkmark$ & 0.47\\
376 & 3 & $03^{\rmn h} 20^{\rmn m} 31\fs809$ & $+00\degr 30\arcmin 28\farcs95$ & 20.80 & 0.3868 & 4 & \nodata & $\checkmark$ & 0.47\\
421 & 2 & $03^{\rmn h} 20^{\rmn m} 33\fs887$ & $+00\degr 30\arcmin 42\farcs03$ & 21.36 & 0.3293 & 4 & $\checkmark$ & \nodata & \nodata\\
271 & 1 & $03^{\rmn h} 20^{\rmn m} 27\fs617$ & $+00\degr 33\arcmin 34\farcs10$ & 20.30 & 0.3751 & 4 & \nodata & $\checkmark$ & 0.55\\
272 & 2 & $03^{\rmn h} 20^{\rmn m} 27\fs678$ & $+00\degr 33\arcmin 35\farcs89$ & 20.21 & 0.3750 & 4 & \nodata & $\checkmark$ & 0.56\\
375 & 2 & $03^{\rmn h} 20^{\rmn m} 31\fs666$ & $+00\degr 30\arcmin 03\farcs86$ & 20.37 & 0.1947 & 4 & $\checkmark$ & \nodata & \nodata\\
387 & 3 & $03^{\rmn h} 20^{\rmn m} 32\fs391$ & $+00\degr 30\arcmin 03\farcs50$ & 20.12 & 0.3807 & 4 & \nodata & $\checkmark$ & 0.61\\
14 & S & $03^{\rmn h} 20^{\rmn m} 23\fs826$ & $+00\degr 33\arcmin 23\farcs52$ & 20.00 & 0.3853 & 4 & \nodata & $\checkmark$ & 0.66\\
262 & 2 & $03^{\rmn h} 20^{\rmn m} 27\fs178$ & $+00\degr 33\arcmin 55\farcs14$ & 21.92 & 0.4733 & 3 & $\checkmark$ & \nodata & \nodata\\
329 & 1 & $03^{\rmn h} 20^{\rmn m} 29\fs739$ & $+00\degr 29\arcmin 38\farcs15$ & 21.15 & 0.3848 & 4 & \nodata & $\checkmark$ & 0.71\\
351 & 2 & $03^{\rmn h} 20^{\rmn m} 30\fs469$ & $+00\degr 29\arcmin 35\farcs85$ & 21.94 & 0.3826 & 4 & $\checkmark$ & $\checkmark$ & 0.72\\
419 & 3 & $03^{\rmn h} 20^{\rmn m} 33\fs764$ & $+00\degr 29\arcmin 47\farcs60$ & 22.29 & 0.3736 & 4 & $\checkmark$ & $\checkmark$ & 0.73\\
249 & 2 & $03^{\rmn h} 20^{\rmn m} 26\fs460$ & $+00\degr 29\arcmin 21\farcs87$ & 22.22 & 0.3260 & 4 & $\checkmark$ & \nodata & \nodata\\
348 & 1 & $03^{\rmn h} 20^{\rmn m} 30\fs356$ & $+00\degr 34\arcmin 47\farcs85$ & 20.64 & 0.3936 & 4 & \nodata & $\checkmark$ & 0.91\\
19 & S & $03^{\rmn h} 20^{\rmn m} 17\fs731$ & $+00\degr 31\arcmin 49\farcs72$ & 19.30 & 0.3851 & 4 & \nodata & $\checkmark$ & 0.94\\
4 & S & $03^{\rmn h} 20^{\rmn m} 27\fs081$ & $+00\degr 28\arcmin 57\farcs99$ & 20.01 & 0.3905 & 4 & \nodata & $\checkmark$ & 0.94\\
427 & 3 & $03^{\rmn h} 20^{\rmn m} 34\fs181$ & $+00\degr 34\arcmin 56\farcs24$ & 19.36 & 0.3250 & 4 & \nodata & \nodata & \nodata\\
364 & 1 & $03^{\rmn h} 20^{\rmn m} 31\fs016$ & $+00\degr 28\arcmin 32\farcs55$ & 19.95 & 0.3911 & 4 & \nodata & \nodata & \nodata\\
98 & 1 & $03^{\rmn h} 20^{\rmn m} 16\fs227$ & $+00\degr 32\arcmin 05\farcs75$ & 21.33 & 0.3690 & 4 & \nodata & \nodata & \nodata\\
236 & 2 & $03^{\rmn h} 20^{\rmn m} 25\fs752$ & $+00\degr 28\arcmin 40\farcs00$ & 20.47 & 0.3896 & 4 & \nodata & \nodata & \nodata\\
246 & 2 & $03^{\rmn h} 20^{\rmn m} 26\fs385$ & $+00\degr 28\arcmin 21\farcs69$ & 22.24 & 0.3929 & 4 & $\checkmark$ & \nodata & \nodata\\
199 & 1 & $03^{\rmn h} 20^{\rmn m} 23\fs492$ & $+00\degr 35\arcmin 13\farcs97$ & 21.79 & 0.4833 & 4 & $\checkmark$ & \nodata & \nodata\\
411 & 2 & $03^{\rmn h} 20^{\rmn m} 33\fs504$ & $+00\degr 35\arcmin 32\farcs84$ & 18.59 & 0.1830 & 4 & \nodata & \nodata & \nodata\\
252 & 3 & $03^{\rmn h} 20^{\rmn m} 26\fs596$ & $+00\degr 35\arcmin 39\farcs38$ & 18.38 & 0.1958 & 4 & \nodata & \nodata & \nodata\\
\hline
\end{tabular}
\end{table*}

In this Section, we describe our measurements of cluster properties: redshift, velocity dispersion, and 
dynamical mass. Throughout we used only galaxies with secure redshifts ($Q \geq 3$). Where needed, we adopt 
the coordinates of the BCG (as listed in \citealt{Menanteau_2013}) as the cluster centre.

\subsection{Cluster redshift measurements}
\label{cluster_red_measurements}
We used the biweight location \citep{Beers_1990} to estimate cluster redshifts. Firstly, we remove obvious 
foreground and background galaxies not physically associated with the cluster by applying a 
3000\,km\,s$^{-1}$ cut relative to the initial cluster redshift (as listed in \citealt{Menanteau_2013}) and 
removed any galaxies determined to be interlopers (see Section~\ref{s_membership} below). We then 
calculated the biweight location from the remaining galaxies. This procedure was iterated until the estimate
for the redshift of the cluster converged. Peculiar velocities for galaxies were then calculated relative to
this newly adopted cluster redshift estimate.

\subsection{Determining cluster membership}
\label{s_membership}
Not all of the galaxies targeted in the SALT RSS field of view are identified as cluster members. For 
this work, we used an adaptation of the fixed-gap method to identify cluster members. This is similar to the 
procedure used by \citet{Fadda_1996} and further refined in \citet{Crawford_2014}. We define the 
peculiar velocity of a galaxy within a cluster as 
\begin{equation}
\Delta v_{i} = c\frac{(z_{i}-\bar{z})}{(1+\bar{z})}\,,
\end{equation}
where $\Delta v_{i}$ is the peculiar velocity of the $i$th galaxy, $z_{i}$ is its redshift, and 
$\bar{z}$ is the redshift of the cluster as estimated using the biweight location 
(see Section \ref{cluster_red_measurements} above). 

To find the 
interlopers, we sorted all galaxies by their peculiar velocities and identified any 
adjacent galaxies (in velocity space) with gaps greater than 1000\,km\,s$^{-1}$. 
\cite{DePropris_2002} argue that galaxy clusters correspond to well-defined peaks with respect to
recessional velocity and that gaps between successive galaxies of more than 1000\,km\,s$^{-1}$ indicate 
interlopers. We iteratively remove galaxies with gaps of greater than 1000\,km\,s$^{-1}$ 
compared to their neighbour until the number of galaxies in the cluster remains constant. 
Any galaxies with projected distance from the cluster centre coordinates greater than $R_{200c}$ 
(the radius within which the mean density is 200 times the critical density of the Universe), were not 
considered to be associated with the cluster and therefore rejected. Note that we relate velocity 
dispersion to cluster mass $M_{200c}$ using a scaling relation, and calculate $R_{200c}$ accordingly
(assuming spherical symmetry; see Section~\ref{cluster_properties} below).

Galaxies flagged as members of 
ACT-CL~J0320.4+0032 are indicated in Table~\ref{t_members_J0320.4+0032}; equivalent tables for the 
other clusters targeted in our SALT observations can be found in the Appendix. 

Over all masks, 47 per cent of 
all successful ($Q \geq 3$) redshift measurements were of galaxies identified as cluster members by the above
procedure.

\subsection{Determining velocity dispersion and mass}
\label{cluster_properties}

\begin{table*}
\caption{Velocity dispersions and derived mass estimates (see Section~\ref{cluster_properties}) for 
ACT clusters observed with SALT. The quantities $M_{500c}$, $R_{500c}$ have been rescaled from 
$M_{200c}$, $R_{200c}$ assuming the appropriate relation of \citet{Duffy_2008}. 
The Members column gives the total number of members for each cluster; the number of square brackets is 
the number of these members with redshifts from SDSS DR10. The $Y_{500c}$ values are rescaled from the values 
in \citet{Hasselfield_2013} for consistency with $R_{500c}$ as determined from the dynamical mass. 
We do not report a dynamical mass measurement for ACT-CL~J0156.4-0123 as only five
members were identified.}
\begin{tabular}{|c|c|c|c|c|c|c|c|c|}
\hline
\label{t_velocityDispersion}
Cluster ID & Members [DR10] & $z$ & $\sigma_{\rm v}$  & $R_{200c}$ & $M_{200c}$             & $R_{500c}$ & $M_{500c}$             & $Y_{500c}$\\
           &                &     & (km\,s$^{-1}$)    & (Mpc)      & (10$^{14}$ M$_{\sun}$) & (Mpc)      & (10$^{14}$ M$_{\sun}$) & (10$^{-4}$ arcmin$^{2}$)\\
\hline
ACT-CL J0320.4$+$0032 & 27 [5]           & 0.3838 & 1430 $\pm$ 160           & 2.3 & 20.0 $\pm$ 5.7           & 1.4 & 12.7 $\pm$ 3.6 & 5.7 $\pm$ 2.0\\
ACT-CL J0348.6$-$0028 & 22 [0]           & 0.3451 & \phantom{0}870 $\pm$ 160 & 1.5 & \phantom{0}5.1 $\pm$ 1.8 & 0.9 & \phantom{0}3.4 $\pm$ 1.2 & 2.6 $\pm$ 1.2\\
ACT-CL J0342.7$-$0017 & 16 [7]           & 0.3069 & 1060 $\pm$ 170           & 1.8 & \phantom{0}9.1 $\pm$ 3.6 & 1.2 & \phantom{0}5.9 $\pm$ 2.3 & 4.3 $\pm$ 2.0\\
ACT-CL J2058.8$+$0123 & 14 [0]           & 0.3273 & \phantom{0}940 $\pm$ 120 & 1.6 & \phantom{0}6.4 $\pm$ 2.1 & 1.0 & \phantom{0}4.2 $\pm$ 1.4 & 6.4 $\pm$ 2.2\\
ACT-CL J0219.9$+$0129 & 13 [5]           & 0.3655 & \phantom{0}900 $\pm$ 210 & 1.5 & \phantom{0}5.6 $\pm$ 3.9 & 1.0 & \phantom{0}3.7 $\pm$ 2.6 & 2.6 $\pm$ 1.9\\
ACT-CL J0045.2$-$0152 & 13 [4]           & 0.5492 & 1020 $\pm$ 250           & 1.5 & \phantom{0}7.2 $\pm$ 4.2 & 1.0 & \phantom{0}4.7 $\pm$ 2.7 & 4.1 $\pm$ 2.5\\ 
ACT-CL J0156.4$-$0123 & \phantom{0}5 [1] & 0.4559 & \nodata                  & \nodata & \nodata                  & \nodata & \nodata          & \nodata \\
\hline
\end{tabular}
\end{table*}

We used the biweight scale estimator 
\citep[described in][]{Beers_1990} to calculate the cluster velocity dispersion $\sigma_v$ from the galaxies selected as members.
Similarly to \citet{Sifon_2013}, we convert our velocity dispersion measurements into estimates of dynamical 
mass by applying a scaling relation measured in cosmological simulations. \citet{Sifon_2013} used the 
relation of \citet{Evrard_2008}, derived from dark matter only simulations, for this purpose. This assumes
that galaxy velocities follow the same relation as dark matter particles in $N$-body simulations. However, 
it has been shown \citep[e.g.,][]{Carlberg_1994, Colin_2000} that the velocity of subhalos is biased 
with respect to the dark matter. So, instead we adopt the relation of \citet{Munari_2013}, which was
calibrated using subhalos and galaxies,
\begin{equation}
\sigma_{\rm v} \, {\rm (km \, s^{-1})} = A \left( \frac{0.7 \times E(z) M_{200}}{10^{15}\, {\rm M}_{\sun}} \right)^{\alpha} ,
\label{e_Munari}
\end{equation}
where $A = (1177 \pm 4.2)$\,km\,s$^{-1}$, $\alpha = 0.364 \pm 0.0021$ and 
$E(z) = \sqrt{\Omega_{\rm m}(1+z)^{3}+\Omega_{\Lambda}}$ (the factor of 0.7 accounts for the assumption of
$H_0 = 70$\,km\,s$^{-1}$\,Mpc$^{-1}$ in this work). The parameters $A$ and $\alpha$ are the normalisation
and slope of the relation \citet{Munari_2013} obtained from a cosmological hydrodynamical simulation including
a model for AGN feedback, using galaxies (with stellar masses $> 3 \times 10^9$\,M$_{\sun}$) as the velocity
tracers (see their Table~1). In comparison to the \citet{Evrard_2008} relation used in \citet{Sifon_2013}, equation (\ref{e_Munari}) results
in masses which are $16-26$ per cent smaller for a given velocity dispersion. This is due to dynamical friction
and tidal disruption and mergers, which act on galaxies but not on dark matter particles \citep{Munari_2013}.
This issue will be discussed in detail in the context of the ACT sample, with reference to numerical simulations,
in Sif\'on et al., in preparation. 

For convenience, and comparison with other results, we convert our $M_{200c}$ estimates into $M_{500c}$, 
following the appropriate $c-M$ relation given in \citet{Duffy_2008}. We estimated uncertainties on all 
cluster properties by bootstrap resampling 5000 times. 

\subsection{Cluster properties}
Table~\ref{t_velocityDispersion} lists the properties we have measured for each cluster, i.e., number of 
members, redshift $z$, velocity dispersion $\sigma_v$, dynamical mass $M_{200c}$ ($M_{500c}$) and associated 
radius $R_{200c}$ ($R_{500c}$), and SZ Comptonisation parameter $Y_{500c}$. The latter have been rescaled from the 
values listed in \citet{Hasselfield_2013} to $R_{500c}$ as determined from the dynamical masses, and we
also account for the fractional error on the dynamical mass in the quoted uncertainty on these rescaled
$Y_{500c}$ measurements. The clusters range in $M_{200c}$ from 
$(5.1 - 20.0) \times 10^{14}$\,M$_{\sun}$ and span the redshift range $0.3 < z < 0.55$. We report spectroscopic
redshifts for the first time in the cases of ACT-CL~J0156.4-0123 and ACT-CL~J2058.8+0123. These new redshift
measurements are in excellent agreement with the photometric redshift estimates for these systems recorded
in \citet{Menanteau_2013}. Note that we do not
report a velocity dispersion measurement and dynamical mass for J0156.4-0123, as only 5 spectroscopic members
were identified in our observations. 
 
\section{Discussion}
\label{s_discussion}

\begin{figure*}
\includegraphics[width=13cm]{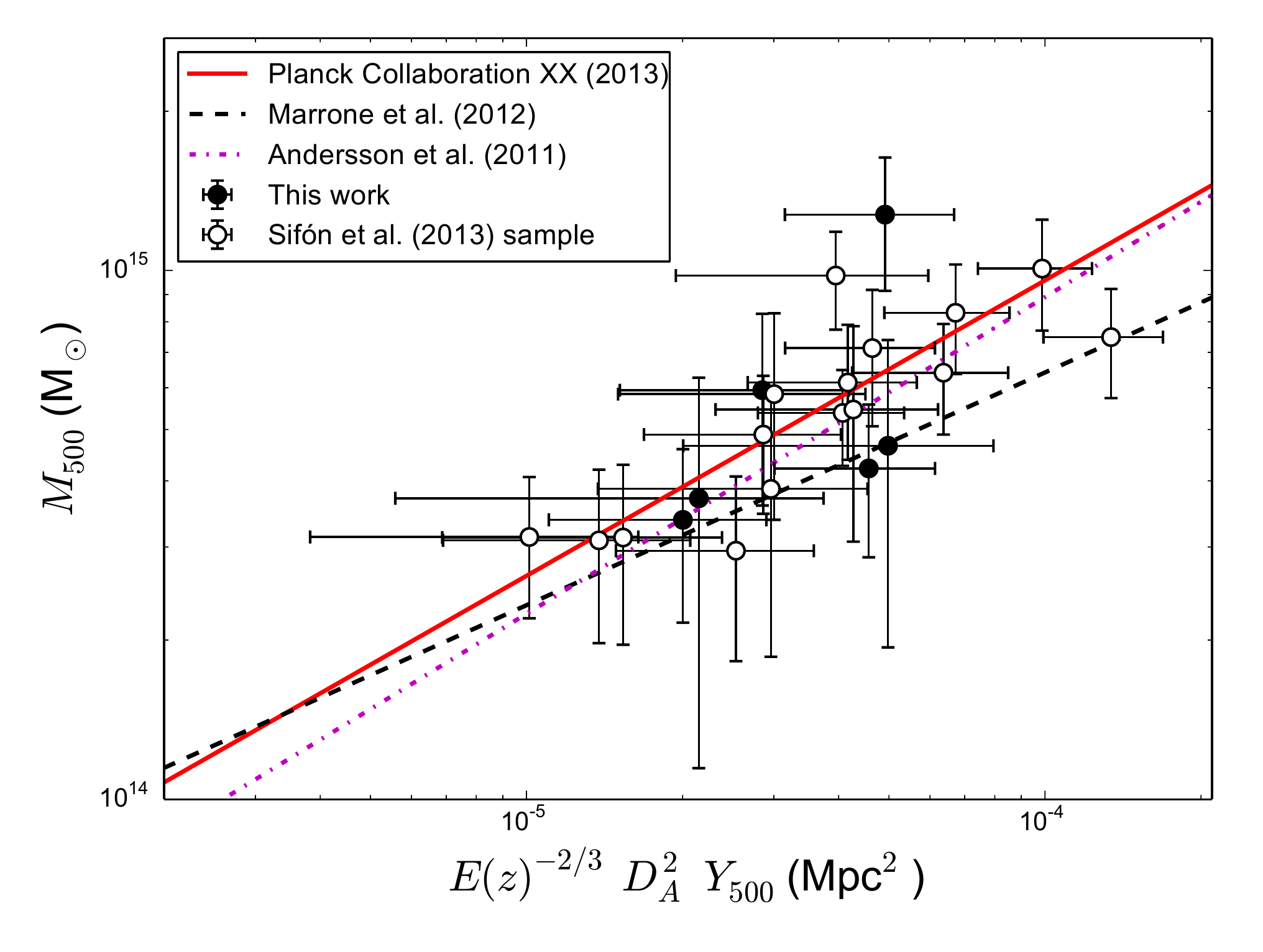}
\caption{Comparison of SALT-derived dynamical masses for ACT equatorial clusters 
(labelled `this work') and the sample of \citet{Sifon_2013}, who obtained such measurements for the ACT 
southern sample \citep{Menanteau_2010, Marriage_2011}. Here all masses (including those of clusters in the 
\citealt{Sifon_2013} sample) have been obtained from velocity
dispersions using the \citet{Munari_2013} scaling relation, rather than the \citet{Evrard_2008} scaling
relation assumed in \citet{Sifon_2013}. All $Y_{500c}$ measurements shown have been rescaled to apertures consistent
with $R_{500c}$ determined from the dynamical masses. We also show recent $Y_{500c} - M_{500c}$ scaling
relations from the literature for comparison: the solid line shows the fiducial relation adopted in 
\citet{Planck_XX_2013} with hydrostatic mass bias $(1-b) = 0.8$ (here the masses are derived from X-ray 
observations); the dashed line shows the relation measured by \citet{Marrone_2012} from SZA
observations of LoCuSS clusters at $z \approx 0.2$, where the masses are measured from weak gravitational 
lensing; and the dot-dashed line shows the relation of \citet{Andersson_2011}, derived from X-ray 
observations of SPT clusters \citep{Vanderlinde_2010}. The SALT dynamical masses appear to be drawn from the 
same distribution as \citet{Sifon_2013}. With the adoption of the \citet{Munari_2013} scaling relation
for conversion of velocity dispersion into mass, we see the ACT clusters are consistent with the 
\citet{Planck_XX_2013} and \citet{Andersson_2011} $Y_{500c}-M_{500c}$ relations, though have a higher
normalisation than the \citet{Marrone_2012} relation.} 
\label{f_YMRelation}
\end{figure*}

\subsection{Previous measurements of the SZ $Y$--mass relation}

As noted in Section~\ref{cluster_properties}, in deriving dynamical mass estimates of the clusters observed
with SALT we have followed the approach of \citet{Sifon_2013}, who observed 16 southern ACT clusters with 
Gemini and the VLT. However, in this work we have adopted the scaling relation of \citet{Munari_2013},
rather than \citet{Evrard_2008}, for the conversion of velocity dispersion into mass. We present a 
comparison of the SALT clusters to the \citet{Sifon_2013} sample in the $Y_{500c} - M_{500c}$ plane 
in Fig.~\ref{f_YMRelation}; note the $M_{200c}$ measurements for all clusters have been converted to 
$M_{500c}$ using the $c-M$ relation of \citet{Duffy_2008}, and the $Y_{500c}$ measurements have been rescaled
from those reported in \citet{Hasselfield_2013} to $R_{500c}$ as determined from the dynamical masses.

As can be seen in Fig.~\ref{f_YMRelation}, the ACT clusters observed with SALT occupy the same region 
of the $Y_{500c}-M_{500c}$ plot as the \citet{Sifon_2013} sample, after converting the velocity dispersions
reported in \citet{Sifon_2013} to masses using the \citet{Munari_2013} scaling relation and rescaling the 
$Y_{500c}$ measurements appropriately. Also 
plotted in Fig.~\ref{f_YMRelation} are some recent $Y_{500c}-M_{500c}$ relations from the literature: the baseline
mass calibration adopted in the \citet{Planck_XX_2013} cosmological study (calibrated from X-ray observations,
and here we assume the hydrostatic bias parameter $b = 0.2$); the relation of \citet{Marrone_2012},
derived from Sunyaev-Zel'dovich Array (SZA) observations of Local Cluster Substructure Survey (LoCuSS) 
clusters, which have mass estimates from gravitational weak lensing \citep{Okabe_2010}; and the 
relation of \citet{Andersson_2011}, with masses measured from \textit{Chandra} and \textit{XMM-Newton} 
observations of South Pole Telescope clusters \citep{Vanderlinde_2010}.

We find that with the adoption of the \citet{Munari_2013} scaling relation, the ACT clusters scatter around the relations
measured by \citet{Planck_XX_2013} and \citet{Andersson_2011}, which are both derived from X-ray observations
(note however that in this work we do not correct for Malmquist-like flux bias as was done in \citealt{Sifon_2013};
the size of this correction is less than 10 per cent for most clusters in the \citealt{Sifon_2013} sample, rising to
20 per cent in the case of the lowest mass cluster, ACT-CL~J0509-5341).
The data have a higher normalisation than is found in the weak-lensing based SZA/LoCuSS measurement
\citep{Marrone_2012}. If we had instead used the \citet{Evrard_2008} $\sigma_v - M_{200}$ scaling, the 
dynamical mass measurements would be $16 - 26$ per cent higher, causing the majority of the ACT clusters 
to lie above all of these recent scaling relation measurements. Such a bias in the scaling relation 
normalisation would lead to larger inferred values for $\sigma_8$ (the normalisation of the dark matter
power spectrum) and $\Omega_{\rm m}$ in a cosmological analysis 
\citep[see, e.g., the discussion in][where the impact of various different scaling relation assumptions
is considered]{Hasselfield_2013}. This issue will be discussed in detail, with reference to results from
cosmological simulations, in Sif\'on et al. (in prep.), 
which will present an updated fit to the $Y_{500c}-M_{500c}$ relation using the full sample of 
48 ACT clusters with velocity dispersion measurements from the literature, Gemini, SALT, and the VLT.

The cluster which deviates most from the southern ACT sample is ACT-CL~J0320.4+0032 (the most massive cluster in this study), 
which has a somewhat lower $Y_{500c}$ than expected given its mass. Based on the uncertainty in its dynamical mass,
it deviates from the \citet{Planck_XX_2013} $Y_{500c}-M_{500c}$ scaling relation by 1.8$\sigma$. If this has
a physical (rather than statistical) cause, it could be due to substructure in the line of
sight velocity distribution; this could lead to an overestimate of the velocity dispersion, and in turn
the dynamical mass. More spectroscopic members need to be identified in order to test if this is the case.
Alternatively, we know that the BCG of this cluster is a quasar host, and it may be possible that recent 
AGN activity has had some influence on the intracluster medium (ICM), and hence the SZ signal, although more data 
are needed to investigate this.

\subsection{ACT-CL J0320.4+0032: a Type II quasar hosted in a Brightest Cluster Galaxy}
\label{s_QSO}
As seen in Fig.~\ref{f_J0320ImageAndSpectra} and noted in Section~\ref{s_redshifts}, the BCG in 
ACT-CL~J0320.4+0032 has relatively broad emission lines, indicating AGN activity. This object has previously been 
identified as a candidate Type II quasar (i.e., an obscured AGN) in the catalogue of \citet{Zakamska_2003}, on the basis of the 
equivalent width of the \textsc{[Oiii]}\,$\lambda$\,5007 line in its SDSS spectrum, and was
subsequently observed with the \textit{Hubble Space Telescope} in November 2006 (PI: H. Schmitt, 
HST Proposal 10880). \citet{VillarMartin_2012} conducted a study of the morphologies of Type II AGN 
hosts using these data, finding that the host galaxy (SDSS~J032029.78+003153.5 in their catalogue)
is an elliptical with a somewhat disturbed morphology, and possibly a double nucleus. \citet{Ramos_2013} identified
this object as being in a clustered environment, but did not note that the host galaxy is actually the BCG of
a massive cluster. The object is not detected as a 1.4\,GHz source in either FIRST \citep[Faint Images of the Radio Sky at Twenty-cm;][]{Becker_1995} or 
NVSS \citep[National Radio Astronomy Observatory Very Large Array Sky Survey;][]{Condon_1998}.

The active BCG of ACT-CL~J0320.4+0032 is a rare discovery, since only a handful of BCGs are known to host 
Type II quasars. The other examples are IRAS~09104+4109 \citep{Kleinmann_1988, OSullivan_2012}, 
Cygnus~A \citep{Antonucci_1994}, Zw8029 \citep{Russell_2013}, and the recent discovery that the central galaxy of 
the Phoenix Cluster at $z = 0.596$ is a Type II quasar \citep{Ueda_2013}. In the latter case, the quasar has
evidently not yet stemmed the cooling of gas, as the central galaxy is also undergoing a starburst 
\citep{McDonald_2012, McDonald_2014}. In addition, some other BCG quasar hosts (NGC~1275, 4C+55.16, 
1H1821+644), all located in cool core clusters, have similar line ratios to ACT-CL~J0320.4+0032,
although they are not formally classified as Type II quasars (A.~Edge, private communication).

The study of such rare systems is important for quantifying the effect of quasar-mode feedback on the 
ICM \citep[see the review by][]{Fabian_2012}. It is well established that radio jets, triggered by radiatively inefficient,
low levels of accretion onto supermassive black holes in BCGs, carve out cavities in the ICM
\citep[e.g.,][]{McNamara_2005, Hlavacek-Larrondo_2013}; indeed, this is the main evidence we have for the
influence of AGN activity on large scales. The gas that fuels radio mode AGN is thought to 
originate in the hot intracluster material, as supported by recent analyses indicating
that radio AGN inhabit environments that support hot atmospheres \citep{Gralla_2014}.
Quasar-mode feedback, on the other hand, is 
radiatively efficient, associated with high accretion rates, drives ubiquitous winds (with velocities 
$\sim 800$\,km\,s$^{-1}$; \citealt{McElroy_2014}), and is thought to be responsible for the quenching
of star formation in massive galaxies \citep[e.g.,][]{DiMatteo_2005, Croton_2006, Bower_2006}. Evolution is 
expected from the quasar-mode to radio-mode \citep[e.g.,][]{Churazov_2005}, with the former including a 
highly obscured stage that keeps the quasar hidden from view in the optical \citep[e.g.,][]{Hopkins_2005}.
Sometimes, radio-emitting bubbles are seen in association with Type II quasars, as in the case of the Teacup AGN
\citep{Harrison_2014}.

Therefore, with only a couple of other similar systems known, ACT-CL~J0320.4+0032 may be an important system to 
study, in order to understand the evolution between these modes of feedback in very massive haloes. As noted
above, the SZ signal for ACT-CL~J0320.4+0032 is relatively low given its dynamical mass, although only at the 
1.8$\sigma$ level. In a study investigating radio-mode feedback, \citet{Gralla_2014} found that the 
SZ effect associated with radio AGN host haloes is somewhat lower than expected from SZ-mass scaling 
relations. The possibility of suppression of the SZ signal by AGN feedback in this cluster (perhaps from
previous radio-mode feedback episodes) could be investigated using
X-ray observations (there are no data on this object in the \textit{Chandra} or \textit{XMM-Newton} archives), 
through measuring the cluster mass with X-ray proxies, and searching for evidence of 
cavities in the X-ray emission. If seen, this would indicate a previous radio-mode feedback episode. With regards to
other Type II quasars hosted in cluster BCGs, we note that
some evidence for cavities has recently been reported on the basis of \textit{Chandra} observations of the Phoenix cluster 
\citep{Hlavacek-Larrondo_2014}, but no cavities have yet been identified in IRAS~09104+4109 \citep{Hlavacek-Larrondo_2013}.
In performing
such a study, care must be taken to separate the emission of the quasar from the 
cluster signal. Such observations, when combined with optical spectroscopy, can also be used to measure the 
obscuration of the nucleus \citep[e.g.,][]{Jia_2013}. Spatially resolved spectroscopic observations may also
be used to investigate outflows from the quasar \citep[e.g.,][]{VillarMartin_2012, McDonald_2014, McElroy_2014}.



%
%
%
%
%

\section{Summary}
\label{s_conclusions}
We have conducted a pilot program of spectroscopic follow-up observations of galaxy clusters discovered via the 
SZ effect, by ACT in its equatorial strip survey, using the RSS instrument on SALT. We successfully 
measured secure redshifts for 191 out of 372 galaxies (overall 51 per cent efficiency) in 7 cluster fields, 
targeting galaxies with $r$-band magnitudes in the range 17.9--23.9, with between 1950--3900\,sec of 
exposure time. 

We measured the redshifts, velocity dispersions, and estimated dynamical masses of the clusters. We made
the first spectroscopic redshift measurements for two systems, ACT-CL~J0156.4-0123 ($z = 0.456$) and 
ACT-CL~J2058.8+0123 ($z = 0.327$), finding these to be in excellent agreement with the photometric
redshift estimates presented in \citet{Menanteau_2013}. Using a scaling relation from the cosmological
hydrodynamical simulations of \citet{Munari_2013} to convert velocity dispersion into mass, we found that 
the clusters range in mass ($M_{200c}$) from $(5-20) \times 10^{14}$\,M$_{\sun}$. The previous study
of ACT cluster dynamical masses \citep{Sifon_2013}, used the \citet{Evrard_2008} scaling relation, based
on the results of dark matter only simulations, to convert velocity dispersion into mass. The 
\citet{Munari_2013}-based masses are $16-26$ per cent smaller. We found that the SALT clusters occupy a 
similar region of the $Y_{500c} - M_{500c}$ plane to the \citet{Sifon_2013} sample, and that 
they are in good agreement with recent measurements of the $Y_{500c}-M_{500c}$ relation measured
based on X-ray observations. The ACT clusters are slightly more massive on average than would be expected
if the \citet{Marrone_2012} weak-lensing based $Y_{500c}-M_{500c}$ relation is used for comparison. A 
future study (Sif\'on et al., in prep.) of the complete sample of 48 ACT clusters with 
dynamical mass measurements from Gemini, SALT, and the VLT will present an updated measurement of the
$Y_{500c}-M_{500c}$ relation, and consider in detail the potential sources of bias in the observational
measurements through comparison with the results of numerical simulations.

In conducting this study, we also found that the BCG in ACT-CL~J0320.4+0032 is host to a previously 
identified Type II quasar \citep{Zakamska_2003, VillarMartin_2012}. However, these previous studies 
were not aware that this object is located in a massive cluster of galaxies, and it is one of only
a handful of such systems that have been discovered. Further follow-up observations of this object may help to
illuminate the role played by quasar-mode feedback in massive clusters.

Overall, this study has proved to be a successful early use of SALT for extragalactic astronomy. These 
results, as well as continued efforts to improve the telescope and instrument performance, 
justify a more extensive use of SALT in the future for exploring higher $z$ clusters, such as those that 
are being discovered with ACTPol \citep{Naess_2014}.

\section*{Acknowledgments}
We thank the anonymous referee for a number of suggestions that improved the quality of this paper. We thank
Alastair Edge for useful discussions about known BCG quasar hosts.
This work is based in large part on observations obtained with the Southern African Large Telescope (SALT). 
Funding for SALT is provided in part by Rutgers University, a founding member of the SALT consortium.
BK, MHi and KM acknowledge financial support from the National Research Foundation and the University of 
KwaZulu-Natal. This work was supported by the U.S. National 
Science Foundation through awards AST-0408698 and AST-0965625 for the ACT project, as well as awards 
PHY-0855887 and PHY-1214379, along with awards AST-0955810 to AJB and AST-1312380 to AK. Funding was also provided by 
Princeton University, the University of 
Pennsylvania, and a Canada Foundation for Innovation (CFI) award to UBC. ACT operates in the Parque 
Astron\'omico Atacama in northern Chile under the auspices of the Comisi\'on Nacional de Investigaci\'on 
Cient\'ifica y Tecnol\'ogica (CONICYT). Computations were performed on the GPC supercomputer at the SciNet
HPC Consortium. SciNet is funded by the CFI under the auspices of Compute Canada, the Government of Ontario, 
the Ontario Research Fund -- Research Excellence; and the University of Toronto.
Funding for SDSS-III has been provided by the Alfred P. Sloan Foundation, the Participating Institutions, 
the National Science Foundation, and the U.S. Department of Energy Office of Science. 
The SDSS-III web site is \url{http://www.sdss3.org/}. SDSS-III is managed by the Astrophysical 
Research Consortium for the Participating Institutions of the SDSS-III Collaboration (see the SDSS-III
web site for details). 

\bibliographystyle{mn2e}
\bibliography{ACTSALT_R1}


\section*{Appendix}

\renewcommand{\thetable}{A\arabic{table}}
\setcounter{table}{0}

\renewcommand{\thefigure}{A\arabic{figure}}
\setcounter{figure}{0}

The tables below list the spectroscopic redshifts measured with SALT RSS in each ACT cluster field. We also
present a selection of images and spectra in the same style as Fig.~\ref{f_J0320ImageAndSpectra}.

\begin{table*}
\caption{Spectroscopic redshifts of galaxies in the direction of ACT-CL J0045.2-0152 measured using SALT RSS;
see Table~\ref{t_members_J0320.4+0032} for an explanation of the table columns.}
\label{t_members_J0045.2-0152}
\begin{tabular}{|c|c|c|c|c|c|c|c|c|c|}
\hline
ID & Mask & RA (J2000) & Dec. (J2000) & $m_{r}$ & $z$ & $Q$ & Em. Lines? & Member? & $r$ (Mpc)\\
\hline
374 & 1 & $00^{\rmn h} 45^{\rmn m} 12\fs499$ & $-01\degr 52\arcmin 31\farcs65$ & 19.22 & 0.5486 & 4 & \nodata & $\checkmark$ & 0.00\\
418 & 1 & $00^{\rmn h} 45^{\rmn m} 14\fs607$ & $-01\degr 52\arcmin 42\farcs69$ & 21.98 & 0.5482 & 4 & \nodata & $\checkmark$ & 0.21\\
18 & S & $00^{\rmn h} 45^{\rmn m} 11\fs507$ & $-01\degr 53\arcmin 09\farcs64$ & 20.66 & 0.5404 & 4 & \nodata & $\checkmark$ & 0.26\\
399 & 1 & $00^{\rmn h} 45^{\rmn m} 13\fs586$ & $-01\degr 53\arcmin 20\farcs61$ & 20.75 & 0.5543 & 4 & $\checkmark$ & $\checkmark$ & 0.33\\
446 & 1 & $00^{\rmn h} 45^{\rmn m} 15\fs661$ & $-01\degr 52\arcmin 08\farcs31$ & 21.99 & 0.5504 & 4 & \nodata & $\checkmark$ & 0.34\\
438 & 1 & $00^{\rmn h} 45^{\rmn m} 15\fs359$ & $-01\degr 53\arcmin 09\farcs28$ & 20.66 & 0.5457 & 4 & $\checkmark$ & $\checkmark$ & 0.37\\
306 & 2 & $00^{\rmn h} 45^{\rmn m} 09\fs483$ & $-01\degr 53\arcmin 17\farcs99$ & 21.47 & 0.9898 & 4 & \nodata & \nodata & \nodata\\
440 & 2 & $00^{\rmn h} 45^{\rmn m} 15\fs476$ & $-01\degr 53\arcmin 29\farcs43$ & 21.38 & 0.4880 & 4 & $\checkmark$ & \nodata & \nodata\\
486 & 2 & $00^{\rmn h} 45^{\rmn m} 17\fs187$ & $-01\degr 51\arcmin 57\farcs75$ & 21.66 & 0.5491 & 4 & \nodata & $\checkmark$ & 0.50\\
434 & 2 & $00^{\rmn h} 45^{\rmn m} 15\fs222$ & $-01\degr 51\arcmin 18\farcs97$ & 21.33 & 0.5490 & 3 & \nodata & $\checkmark$ & 0.53\\
7 & S & $00^{\rmn h} 45^{\rmn m} 07\fs935$ & $-01\degr 53\arcmin 21\farcs18$ & 20.40 & 0.5488 & 4 & \nodata & $\checkmark$ & 0.54\\
451 & 1 & $00^{\rmn h} 45^{\rmn m} 15\fs949$ & $-01\degr 53\arcmin 40\farcs05$ & 19.88 & 0.5535 & 4 & \nodata & $\checkmark$ & 0.55\\
376 & 1 & $00^{\rmn h} 45^{\rmn m} 12\fs645$ & $-01\degr 53\arcmin 57\farcs37$ & 21.48 & 0.5574 & 3 & \nodata & $\checkmark$ & 0.55\\
509 & 1 & $00^{\rmn h} 45^{\rmn m} 18\fs228$ & $-01\degr 52\arcmin 19\farcs59$ & 21.93 & 0.7100 & 4 & $\checkmark$ & \nodata & \nodata\\
387 & 1 & $00^{\rmn h} 45^{\rmn m} 13\fs094$ & $-01\degr 51\arcmin 04\farcs19$ & 21.45 & 0.5130 & 4 & $\checkmark$ & \nodata & \nodata\\
335 & 2 & $00^{\rmn h} 45^{\rmn m} 10\fs733$ & $-01\degr 53\arcmin 59\farcs12$ & 21.66 & 0.6379 & 4 & \nodata & \nodata & \nodata\\
14 & S & $00^{\rmn h} 45^{\rmn m} 08\fs649$ & $-01\degr 54\arcmin 07\farcs50$ & 20.77 & 0.5526 & 4 & \nodata & $\checkmark$ & 0.72\\
251 & 2 & $00^{\rmn h} 45^{\rmn m} 06\fs334$ & $-01\degr 53\arcmin 46\farcs27$ & 21.34 & 0.6425 & 3 & $\checkmark$ & \nodata & \nodata\\
11 & S & $00^{\rmn h} 45^{\rmn m} 09\fs245$ & $-01\degr 54\arcmin 26\farcs52$ & 20.04 & 0.5388 & 4 & \nodata & $\checkmark$ & 0.80\\
439 & 1 & $00^{\rmn h} 45^{\rmn m} 15\fs372$ & $-01\degr 50\arcmin 28\farcs54$ & 21.44 & 0.6570 & 4 & $\checkmark$ & \nodata & \nodata\\
400 & 1 & $00^{\rmn h} 45^{\rmn m} 13\fs59$ & $-01\degr 54\arcmin 56\farcs36$ & 20.47 & 0.3675 & 3 & \nodata & \nodata & \nodata\\
415 & 1 & $00^{\rmn h} 45^{\rmn m} 14\fs368$ & $-01\degr 49\arcmin 53\farcs20$ & 20.15 & 0.2431 & 4 & $\checkmark$ & \nodata & \nodata\\
207 & 1 & $00^{\rmn h} 45^{\rmn m} 03\fs524$ & $-01\degr 50\arcmin 51\farcs21$ & 20.18 & 0.4726 & 4 & \nodata & \nodata & \nodata\\
363 & 1 & $00^{\rmn h} 45^{\rmn m} 12\fs009$ & $-01\degr 49\arcmin 37\farcs58$ & 21.18 & 0.5434 & 4 & \nodata & \nodata & \nodata\\
510 & 2 & $00^{\rmn h} 45^{\rmn m} 18\fs243$ & $-01\degr 55\arcmin 20\farcs68$ & 21.61 & 0.5284 & 3 & $\checkmark$ & \nodata & \nodata\\
355 & 1 & $00^{\rmn h} 45^{\rmn m} 11\fs751$ & $-01\degr 56\arcmin 24\farcs72$ & 21.26 & 0.5523 & 4 & $\checkmark$ & \nodata & \nodata\\
343 & 1 & $00^{\rmn h} 45^{\rmn m} 11\fs012$ & $-01\degr 56\arcmin 39\farcs24$ & 21.58 & 0.2010 & 4 & \nodata & \nodata & \nodata\\
371 & 2 & $00^{\rmn h} 45^{\rmn m} 12\fs315$ & $-01\degr 56\arcmin 48\farcs94$ & 19.37 & 0.8268 & 4 & \nodata & \nodata & \nodata\\
\hline
\end{tabular}
\end{table*}

\begin{figure*}
\includegraphics[width=17.5cm]{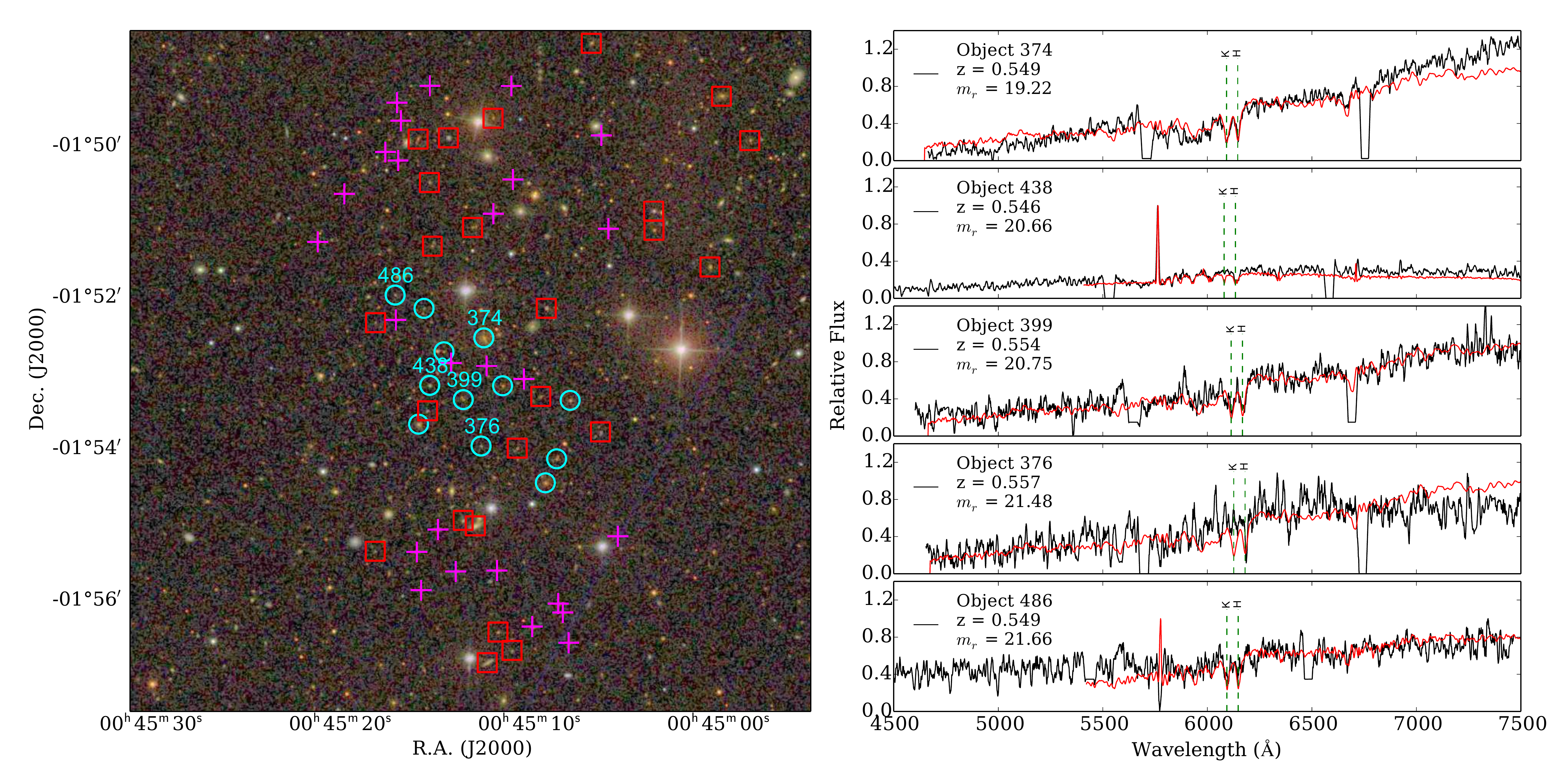}
\caption{The $z = 0.55$ cluster ACT-CL J0045.2-0152 (see Fig.~\ref{f_J0320ImageAndSpectra} for an explanation
of symbols and colours).}
\label{f_J0045ImagesAndSpectra}
\end{figure*}

\begin{table*}
\caption{Spectroscopic redshifts of galaxies in the direction of ACT-CL J0156.4-0123 measured using SALT RSS;
see Table~\ref{t_members_J0320.4+0032} for an explanation of the table columns.}
\label{t_members_J0156.4-0123}
\begin{tabular}{|c|c|c|c|c|c|c|c|c|c|}
\hline
ID & Mask & RA (J2000) & Dec. (J2000) & $m_{r}$ & $z$ & $Q$ & Em. Lines? & Member? & $r$ (Mpc)\\
\hline
320 & 1 & $01^{\rmn h} 56^{\rmn m} 24\fs297$ & $-01\degr 23\arcmin 17\farcs32$ & 17.88 & 0.4526 & 4 & \nodata & $\checkmark$ & 0.00\\
317 & 1 & $01^{\rmn h} 56^{\rmn m} 24\fs192$ & $-01\degr 23\arcmin 35\farcs61$ & 20.34 & 0.4380 & 4 & \nodata & \nodata & \nodata\\
6 & S & $01^{\rmn h} 56^{\rmn m} 26\fs066$ & $-01\degr 23\arcmin 33\farcs76$ & 19.55 & 0.4602 & 4 & \nodata & $\checkmark$ & 0.18\\
246 & 3 & $01^{\rmn h} 56^{\rmn m} 20\fs883$ & $-01\degr 23\arcmin 55\farcs64$ & 21.45 & 0.5965 & 3 & \nodata & \nodata & \nodata\\
268 & 1 & $01^{\rmn h} 56^{\rmn m} 21\fs674$ & $-01\degr 22\arcmin 12\farcs07$ & 21.08 & 0.4435 & 4 & $\checkmark$ & \nodata & \nodata\\
403 & 1 & $01^{\rmn h} 56^{\rmn m} 28\fs970$ & $-01\degr 22\arcmin 46\farcs03$ & 20.22 & 0.4582 & 4 & $\checkmark$ & $\checkmark$ & 0.44\\
239 & 1 & $01^{\rmn h} 56^{\rmn m} 20\fs400$ & $-01\degr 22\arcmin 24\farcs40$ & 20.78 & 0.5689 & 4 & \nodata & \nodata & \nodata\\
310 & 1 & $01^{\rmn h} 56^{\rmn m} 23\fs810$ & $-01\degr 21\arcmin 57\farcs79$ & 19.71 & 0.2769 & 4 & \nodata & \nodata & \nodata\\
410 & 1 & $01^{\rmn h} 56^{\rmn m} 29\fs305$ & $-01\degr 23\arcmin 59\farcs72$ & 20.69 & 0.4561 & 4 & \nodata & $\checkmark$ & 0.50\\
243 & 1 & $01^{\rmn h} 56^{\rmn m} 20\fs696$ & $-01\degr 21\arcmin 30\farcs28$ & 19.90 & 0.5597 & 4 & $\checkmark$ & $\checkmark$ & 0.69\\
195 & 3 & $01^{\rmn h} 56^{\rmn m} 17\fs450$ & $-01\degr 22\arcmin 06\farcs30$ & 21.92 & 0.5985 & 3 & $\checkmark$ & \nodata & \nodata\\
361 & 1 & $01^{\rmn h} 56^{\rmn m} 26\fs472$ & $-01\degr 25\arcmin 33\farcs21$ & 18.22 & 0.1368 & 4 & $\checkmark$ & \nodata & \nodata\\
186 & 3 & $01^{\rmn h} 56^{\rmn m} 16\fs882$ & $-01\degr 21\arcmin 51\farcs12$ & 20.06 & 0.5683 & 4 & \nodata & \nodata & \nodata\\
419 & 1 & $01^{\rmn h} 56^{\rmn m} 30\fs384$ & $-01\degr 21\arcmin 07\farcs92$ & 20.60 & 0.6804 & 4 & $\checkmark$ & \nodata & \nodata\\
164 & 1 & $01^{\rmn h} 56^{\rmn m} 15\fs462$ & $-01\degr 24\arcmin 47\farcs50$ & 20.03 & 0.6058 & 4 & $\checkmark$ & \nodata & \nodata\\
355 & 1 & $01^{\rmn h} 56^{\rmn m} 26\fs243$ & $-01\degr 25\arcmin 57\farcs64$ & 19.66 & 0.3397 & 4 & $\checkmark$ & \nodata & \nodata\\
393 & 1 & $01^{\rmn h} 56^{\rmn m} 28\fs181$ & $-01\degr 20\arcmin 43\farcs36$ & 20.68 & 0.3941 & 4 & \nodata & \nodata & \nodata\\
142 & 3 & $01^{\rmn h} 56^{\rmn m} 13\fs247$ & $-01\degr 22\arcmin 40\farcs53$ & 19.98 & 0.7713 & 3 & $\checkmark$ & \nodata & \nodata\\
350 & 3 & $01^{\rmn h} 56^{\rmn m} 25\fs923$ & $-01\degr 20\arcmin 21\farcs53$ & 18.60 & 0.3808 & 4 & $\checkmark$ & \nodata & \nodata\\
468 & 3 & $01^{\rmn h} 56^{\rmn m} 34\fs261$ & $-01\degr 21\arcmin 38\farcs25$ & 19.58 & 0.3400 & 4 & \nodata & \nodata & \nodata\\
123 & 3 & $01^{\rmn h} 56^{\rmn m} 12\fs040$ & $-01\degr 23\arcmin 23\farcs13$ & 19.12 & 0.4773 & 4 & $\checkmark$ & \nodata & \nodata\\
337 & 1 & $01^{\rmn h} 56^{\rmn m} 25\fs235$ & $-01\degr 20\arcmin 12\farcs52$ & 20.15 & 0.0392 & 4 & \nodata & \nodata & \nodata\\
335 & 1 & $01^{\rmn h} 56^{\rmn m} 25\fs171$ & $-01\degr 26\arcmin 31\farcs22$ & 20.65 & 0.4551 & 4 & \nodata & \nodata & \nodata\\
426 & 1 & $01^{\rmn h} 56^{\rmn m} 30\fs748$ & $-01\degr 26\arcmin 19\farcs24$ & 20.93 & 0.4497 & 4 & \nodata & \nodata & \nodata\\
340 & 1 & $01^{\rmn h} 56^{\rmn m} 25\fs337$ & $-01\degr 26\arcmin 46\farcs89$ & 21.09 & 0.7356 & 4 & $\checkmark$ & \nodata & \nodata\\
385 & 3 & $01^{\rmn h} 56^{\rmn m} 27\fs728$ & $-01\degr 19\arcmin 40\farcs68$ & 21.59 & 0.3845 & 3 & \nodata & \nodata & \nodata\\
127 & 3 & $01^{\rmn h} 56^{\rmn m} 12\fs255$ & $-01\degr 20\arcmin 47\farcs78$ & 19.70 & 0.4215 & 3 & $\checkmark$ & \nodata & \nodata\\
\hline
\end{tabular}
\end{table*}

\begin{figure*}
\includegraphics[width=17.5cm]{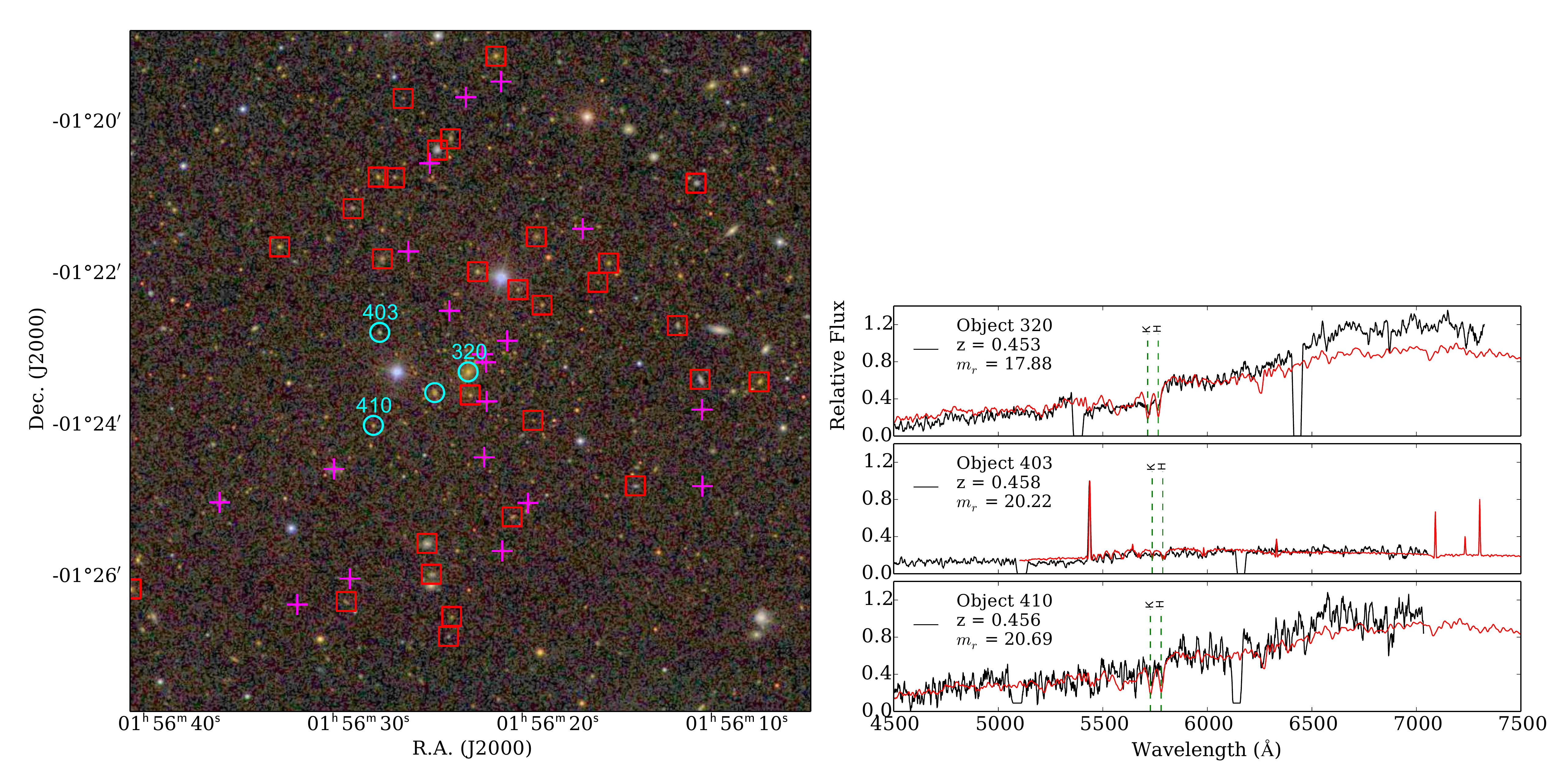}
\caption{The $z$ = 0.46 cluster ACT-CL J0156.4-0123 (see Fig.~\ref{f_J0320ImageAndSpectra} for an explanation
of symbols and colours) The unlabelled member galaxy is from SDSS DR10.}
\label{f_J0156ImagesAndSpectra}
\end{figure*}

\begin{table*}
\caption{Spectroscopic redshifts of galaxies in the direction of ACT-CL J0219.9+0129 measured using SALT RSS;
see Table~\ref{t_members_J0320.4+0032} for an explanation of the table columns.}
\label{t_members_J0219.9+0129}
\begin{tabular}{|c|c|c|c|c|c|c|c|c|c|}
\hline
ID & Mask & RA (J2000) & Dec. (J2000) & $m_{r}$ & $z$ & $Q$ & Em. Lines? & Member? & $r$ (Mpc)\\
\hline
370 & 1 & $02^{\rmn h} 19^{\rmn m} 52\fs155$ & $+01\degr 29\arcmin 52\farcs19$ & 17.96 & 0.3646 & 4 & \nodata & $\checkmark$ & 0.00\\
397 & 1 & $02^{\rmn h} 19^{\rmn m} 52\fs975$ & $+01\degr 29\arcmin 35\farcs03$ & 21.29 & 0.3639 & 4 & \nodata & $\checkmark$ & 0.11\\
410 & 1 & $02^{\rmn h} 19^{\rmn m} 53\fs386$ & $+01\degr 30\arcmin 31\farcs71$ & 21.22 & 0.3679 & 4 & \nodata & $\checkmark$ & 0.22\\
434 & 1 & $02^{\rmn h} 19^{\rmn m} 54\fs222$ & $+01\degr 29\arcmin 20\farcs51$ & 21.01 & 0.3641 & 4 & \nodata & $\checkmark$ & 0.22\\
390 & 1 & $02^{\rmn h} 19^{\rmn m} 52\fs668$ & $+01\degr 29\arcmin 06\farcs20$ & 21.01 & 0.3585 & 4 & \nodata & $\checkmark$ & 0.23\\
1 & S & $02^{\rmn h} 19^{\rmn m} 57\fs414$ & $+01\degr 30\arcmin 02\farcs31$ & 19.45 & 0.3697 & 4 & \nodata & $\checkmark$ & 0.40\\
4 & S & $02^{\rmn h} 19^{\rmn m} 56\fs219$ & $+01\degr 30\arcmin 58\farcs53$ & 18.56 & 0.3675 & 4 & \nodata & $\checkmark$ & 0.45\\
395 & 1 & $02^{\rmn h} 19^{\rmn m} 52\fs931$ & $+01\degr 31\arcmin 22\farcs76$ & 20.08 & 0.3544 & 4 & \nodata & $\checkmark$ & 0.45\\
379 & 1 & $02^{\rmn h} 19^{\rmn m} 52\fs348$ & $+01\degr 28\arcmin 17\farcs92$ & 20.17 & 0.3496 & 4 & \nodata & \nodata & \nodata\\
508 & 1 & $02^{\rmn h} 19^{\rmn m} 55\fs973$ & $+01\degr 31\arcmin 07\farcs71$ & 21.23 & 0.3649 & 4 & \nodata & $\checkmark$ & 0.47\\
309 & 1 & $02^{\rmn h} 19^{\rmn m} 49\fs797$ & $+01\degr 31\arcmin 35\farcs97$ & 20.67 & 0.3489 & 4 & \nodata & \nodata & \nodata\\
407 & 1 & $02^{\rmn h} 19^{\rmn m} 53\fs365$ & $+01\degr 31\arcmin 48\farcs61$ & 19.11 & 0.2389 & 4 & \nodata & \nodata & \nodata\\
447 & 1 & $02^{\rmn h} 19^{\rmn m} 54\fs580$ & $+01\degr 27\arcmin 59\farcs68$ & 20.91 & 0.3686 & 4 & \nodata & $\checkmark$ & 0.59\\
274 & 1 & $02^{\rmn h} 19^{\rmn m} 48\fs609$ & $+01\degr 27\arcmin 46\farcs90$ & 18.53 & 0.3666 & 4 & \nodata & $\checkmark$ & 0.67\\
428 & 1 & $02^{\rmn h} 19^{\rmn m} 53\fs908$ & $+01\degr 32\arcmin 35\farcs19$ & 21.45 & 0.5602 & 4 & \nodata & \nodata & \nodata\\
5 & S & $02^{\rmn h} 20^{\rmn m} 01\fs648$ & $+01\degr 28\arcmin 27\farcs15$ & 19.19 & 0.3629 & 4 & \nodata & $\checkmark$ & 0.82\\
450 & 1 & $02^{\rmn h} 19^{\rmn m} 54\fs615$ & $+01\degr 26\arcmin 53\farcs96$ & 20.06 & 0.3697 & 4 & \nodata & $\checkmark$ & 0.90\\
229 & 1 & $02^{\rmn h} 19^{\rmn m} 46\fs790$ & $+01\degr 26\arcmin 39\farcs57$ & 21.61 & 0.7861 & 4 & $\checkmark$ & \nodata & \nodata\\
299 & 1 & $02^{\rmn h} 19^{\rmn m} 49\fs495$ & $+01\degr 26\arcmin 20\farcs61$ & 20.68 & 0.5314 & 4 & $\checkmark$ & \nodata & \nodata\\
663 & 1 & $02^{\rmn h} 20^{\rmn m} 02\fs912$ & $+01\degr 27\arcmin 08\farcs25$ & 21.69 & 0.3580 & 3 & \nodata & \nodata & \nodata\\
\hline
\end{tabular}
\end{table*}

\begin{figure*}
\includegraphics[width=17.5cm]{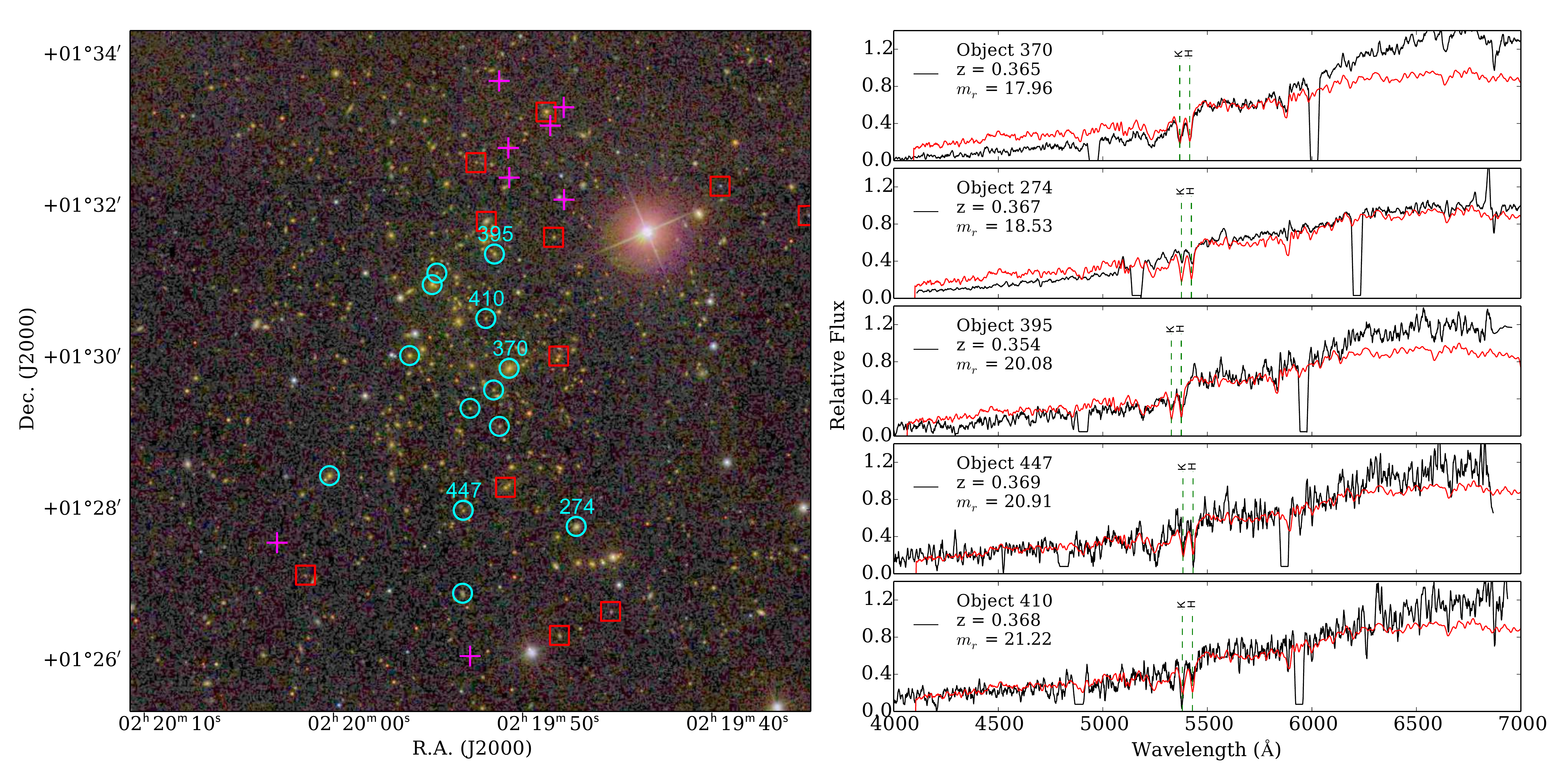}
\caption{The $z = 0.36$ cluster ACT-CL J0219.9+0129 (see Fig.~\ref{f_J0320ImageAndSpectra} for an explanation
of symbols and colours).}
\label{f_J0219ImageAndSpectra}
\end{figure*}

\begin{table*}
\caption{Spectroscopic redshifts of galaxies in the direction of ACT-CL J0342.7-0017 measured using SALT RSS;
see Table~\ref{t_members_J0320.4+0032} for an explanation of the table columns.}
\label{t_members_J0342.7-0017}
\begin{tabular}{|c|c|c|c|c|c|c|c|c|c|}
\hline
ID & Mask & RA (J2000) & Dec. (J2000) & $m_{r}$ & $z$ & $Q$ & Em. Lines? & Member? & $r$ (Mpc)\\
\hline
7 & S & $03^{\rmn h} 42^{\rmn m} 42\fs651$ & $-00\degr 17\arcmin 08\farcs29$ & 17.70 & 0.3072 & 4 & \nodata & $\checkmark$ & 0.00\\
2 & S & $03^{\rmn h} 42^{\rmn m} 42\fs873$ & $-00\degr 17\arcmin 10\farcs22$ & 18.20 & 0.3127 & 4 & \nodata & $\checkmark$ & 0.02\\
362 & 1 & $03^{\rmn h} 42^{\rmn m} 42\fs346$ & $-00\degr 17\arcmin 01\farcs28$ & 20.58 & 0.3052 & 4 & \nodata & $\checkmark$ & 0.04\\
364 & 1 & $03^{\rmn h} 42^{\rmn m} 42\fs414$ & $-00\degr 16\arcmin 42\farcs67$ & 20.21 & 0.3039 & 4 & \nodata & $\checkmark$ & 0.11\\
413 & 1 & $03^{\rmn h} 42^{\rmn m} 44\fs804$ & $-00\degr 17\arcmin 18\farcs92$ & 20.25 & 0.3009 & 4 & \nodata & $\checkmark$ & 0.15\\
432 & 1 & $03^{\rmn h} 42^{\rmn m} 45\fs707$ & $-00\degr 17\arcmin 37\farcs69$ & 20.47 & 0.1658 & 3 & \nodata & \nodata & \nodata\\
374 & 1 & $03^{\rmn h} 42^{\rmn m} 42\fs883$ & $-00\degr 16\arcmin 09\farcs36$ & 20.86 & 0.3010 & 4 & $\checkmark$ & $\checkmark$ & 0.26\\
429 & 1 & $03^{\rmn h} 42^{\rmn m} 45\fs595$ & $-00\degr 16\arcmin 22\farcs35$ & 21.00 & 0.3111 & 4 & \nodata & $\checkmark$ & 0.29\\
8 & S & $03^{\rmn h} 42^{\rmn m} 38\fs364$ & $-00\degr 16\arcmin 45\farcs57$ & 18.67 & 0.3111 & 4 & \nodata & $\checkmark$ & 0.30\\
409 & 1 & $03^{\rmn h} 42^{\rmn m} 44\fs651$ & $-00\degr 18\arcmin 14\farcs94$ & 19.81 & 0.3107 & 4 & \nodata & $\checkmark$ & 0.33\\
363 & 1 & $03^{\rmn h} 42^{\rmn m} 42\fs387$ & $-00\degr 15\arcmin 54\farcs14$ & 20.23 & 0.7064 & 4 & \nodata & \nodata & \nodata\\
25 & S & $03^{\rmn h} 42^{\rmn m} 44\fs899$ & $-00\degr 18\arcmin 22\farcs30$ & 19.86 & 0.3129 & 4 & \nodata & $\checkmark$ & 0.36\\
474 & 1 & $03^{\rmn h} 42^{\rmn m} 47\fs751$ & $-00\degr 17\arcmin 51\farcs14$ & 20.56 & 0.1654 & 4 & \nodata & \nodata & \nodata\\
400 & 1 & $03^{\rmn h} 42^{\rmn m} 44\fs194$ & $-00\degr 15\arcmin 36\farcs13$ & 20.57 & 0.2384 & 4 & $\checkmark$ & \nodata & \nodata\\
407 & 1 & $03^{\rmn h} 42^{\rmn m} 44\fs546$ & $-00\degr 18\arcmin 40\farcs94$ & 20.95 & 0.3019 & 4 & \nodata & $\checkmark$ & 0.43\\
216 & 1 & $03^{\rmn h} 42^{\rmn m} 35\fs633$ & $-00\degr 18\arcmin 02\farcs22$ & 20.77 & 0.3664 & 4 & $\checkmark$ & \nodata & \nodata\\
436 & 1 & $03^{\rmn h} 42^{\rmn m} 45\fs840$ & $-00\degr 15\arcmin 18\farcs69$ & 19.50 & 0.2393 & 4 & \nodata & \nodata & \nodata\\
274 & 1 & $03^{\rmn h} 42^{\rmn m} 38\fs579$ & $-00\degr 15\arcmin 00\farcs56$ & 20.70 & 0.0191 & 3 & \nodata & \nodata & \nodata\\
473 & 1 & $03^{\rmn h} 42^{\rmn m} 47\fs601$ & $-00\degr 19\arcmin 18\farcs95$ & 20.62 & 0.2863 & 4 & $\checkmark$ & \nodata & \nodata\\
26 & S & $03^{\rmn h} 42^{\rmn m} 34\fs639$ & $-00\degr 15\arcmin 29\farcs36$ & 18.95 & 0.3043 & 4 & \nodata & $\checkmark$ & 0.69\\
42 & S & $03^{\rmn h} 42^{\rmn m} 52\fs148$ & $-00\degr 18\arcmin 08\farcs46$ & 18.98 & 0.3042 & 4 & \nodata & $\checkmark$ & 0.69\\
260 & 1 & $03^{\rmn h} 42^{\rmn m} 37\fs754$ & $-00\degr 19\arcmin 30\farcs24$ & 19.15 & 0.3033 & 4 & \nodata & $\checkmark$ & 0.71\\
5 & S & $03^{\rmn h} 42^{\rmn m} 31\fs905$ & $-00\degr 16\arcmin 49\farcs20$ & 18.74 & 0.3113 & 4 & \nodata & $\checkmark$ & 0.72\\
428 & 1 & $03^{\rmn h} 42^{\rmn m} 45\fs590$ & $-00\degr 19\arcmin 50\farcs29$ & 20.61 & 0.1116 & 4 & $\checkmark$ & \nodata & \nodata\\
430 & 1 & $03^{\rmn h} 42^{\rmn m} 45\fs654$ & $-00\degr 20\arcmin 12\farcs98$ & 20.77 & 0.3093 & 4 & \nodata & $\checkmark$ & 0.85\\
468 & 1 & $03^{\rmn h} 42^{\rmn m} 47\fs405$ & $-00\degr 20\arcmin 24\farcs07$ & 19.67 & 0.3656 & 4 & \nodata & \nodata & \nodata\\
498 & 1 & $03^{\rmn h} 42^{\rmn m} 48\fs728$ & $-00\degr 20\arcmin 44\farcs52$ & 20.80 & 0.2869 & 4 & $\checkmark$ & \nodata & \nodata\\
521 & 1 & $03^{\rmn h} 42^{\rmn m} 50\fs300$ & $-00\degr 20\arcmin 58\farcs49$ & 21.30 & 0.4619 & 4 & \nodata & \nodata & \nodata\\
\hline
\end{tabular}
\end{table*}

\begin{figure*}
\includegraphics[width=17.5cm]{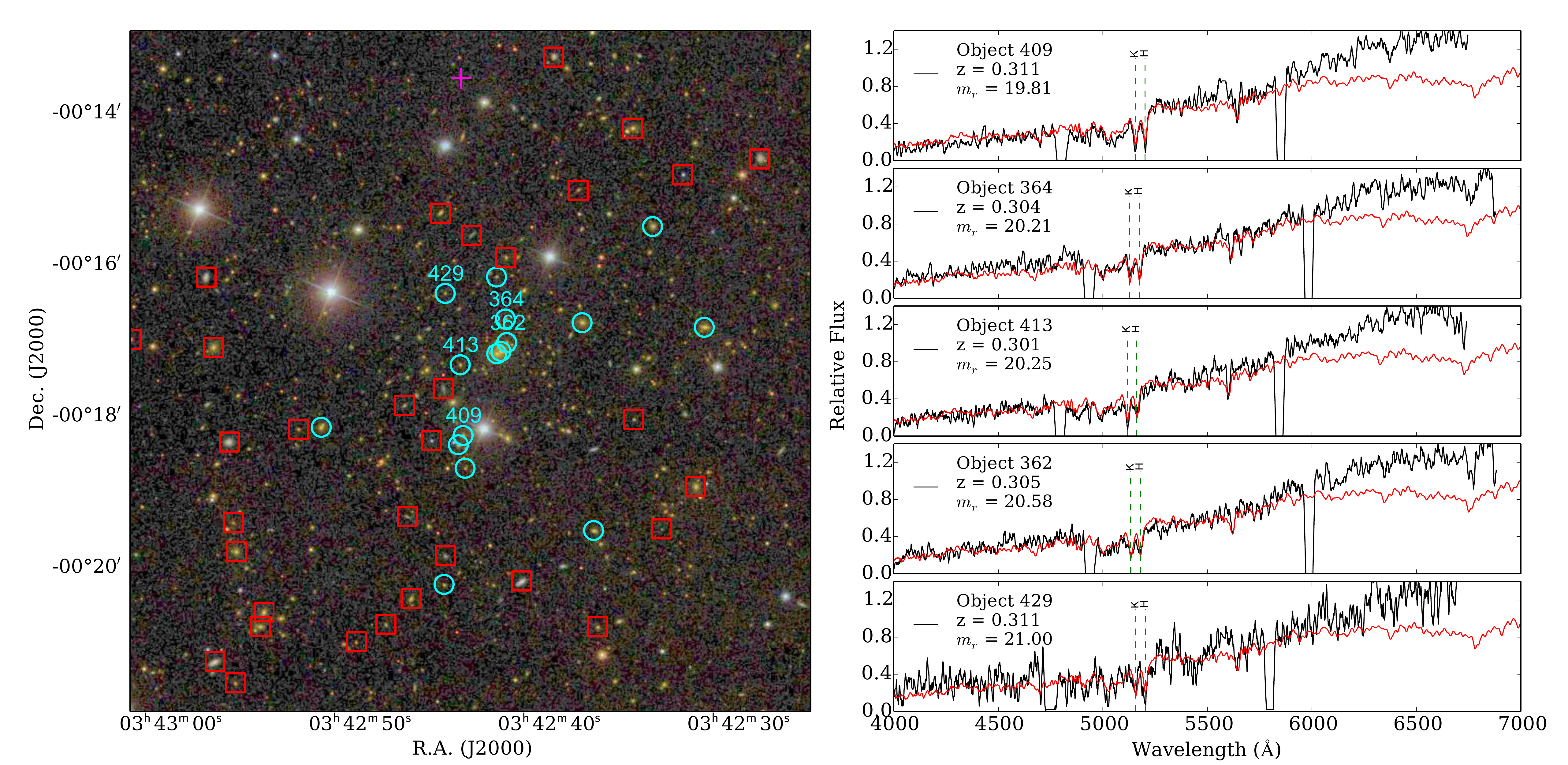}
\caption{The $z = 0.30$ cluster ACT-CL J0342.7-0017 (see Fig.~\ref{f_J0320ImageAndSpectra} for an explanation
of symbols and colours).}
\label{f_J0342ImagesAndSpectra}
\end{figure*}

\begin{table*}
\caption{Spectroscopic redshifts of galaxies in the direction of ACT-CL J0348.6-0028 measured using SALT RSS;
see Table~\ref{t_members_J0320.4+0032} for an explanation of the table columns.}
\label{t_members_J0348.6-0028}
\begin{tabular}{|c|c|c|c|c|c|c|c|c|c|}
\hline
ID & Mask & RA (J2000) & Dec. (J2000) & $m_{r}$ & $z$ & $Q$ & Em. Lines? & Member? & $r$ (Mpc)\\
\hline
274 & 3 & $03^{\rmn h} 48^{\rmn m} 39\fs545$ & $-00\degr 28\arcmin 16\farcs90$ & 19.22 & 0.3450 & 4 & \nodata & $\checkmark$ & 0.00\\
252 & 4 & $03^{\rmn h} 48^{\rmn m} 38\fs726$ & $-00\degr 28\arcmin 07\farcs42$ & 18.13 & 0.1381 & 4 & \nodata & \nodata & \nodata\\
246 & 1 & $03^{\rmn h} 48^{\rmn m} 38\fs378$ & $-00\degr 28\arcmin 23\farcs21$ & 20.49 & 0.3443 & 4 & \nodata & $\checkmark$ & 0.09\\
229 & 4 & $03^{\rmn h} 48^{\rmn m} 37\fs682$ & $-00\degr 28\arcmin 20\farcs77$ & 20.07 & 0.1393 & 4 & \nodata & \nodata & \nodata\\
303 & 1 & $03^{\rmn h} 48^{\rmn m} 41\fs534$ & $-00\degr 28\arcmin 07\farcs50$ & 19.87 & 0.3443 & 4 & \nodata & $\checkmark$ & 0.16\\
230 & 1 & $03^{\rmn h} 48^{\rmn m} 37\fs699$ & $-00\degr 28\arcmin 36\farcs14$ & 20.40 & 0.3416 & 4 & \nodata & $\checkmark$ & 0.16\\
232 & 3 & $03^{\rmn h} 48^{\rmn m} 37\fs909$ & $-00\degr 28\arcmin 46\farcs75$ & 21.04 & 0.3448 & 3 & \nodata & $\checkmark$ & 0.19\\
315 & 3 & $03^{\rmn h} 48^{\rmn m} 42\fs291$ & $-00\degr 28\arcmin 29\farcs17$ & 21.08 & 0.3490 & 3 & \nodata & $\checkmark$ & 0.22\\
218 & 2 & $03^{\rmn h} 48^{\rmn m} 37\fs125$ & $-00\degr 27\arcmin 48\farcs38$ & 18.17 & 0.1603 & 4 & \nodata & \nodata & \nodata\\
264 & 1 & $03^{\rmn h} 48^{\rmn m} 39\fs134$ & $-00\degr 29\arcmin 05\farcs30$ & 21.02 & 0.3503 & 4 & \nodata & $\checkmark$ & 0.24\\
312 & 1 & $03^{\rmn h} 48^{\rmn m} 42\fs151$ & $-00\degr 28\arcmin 48\farcs76$ & 19.76 & 0.3462 & 4 & \nodata & $\checkmark$ & 0.25\\
304 & 2 & $03^{\rmn h} 48^{\rmn m} 41\fs593$ & $-00\degr 27\arcmin 35\farcs15$ & 19.48 & 0.5471 & 4 & \nodata & \nodata & \nodata\\
207 & 1 & $03^{\rmn h} 48^{\rmn m} 36\fs300$ & $-00\degr 27\arcmin 49\farcs82$ & 20.50 & 0.3401 & 4 & \nodata & $\checkmark$ & 0.27\\
308 & 3 & $03^{\rmn h} 48^{\rmn m} 42\fs024$ & $-00\degr 28\arcmin 58\farcs10$ & 21.09 & 0.3420 & 4 & \nodata & $\checkmark$ & 0.28\\
239 & 1 & $03^{\rmn h} 48^{\rmn m} 38\fs081$ & $-00\degr 29\arcmin 18\farcs25$ & 20.68 & 0.3482 & 4 & \nodata & $\checkmark$ & 0.32\\
330 & 3 & $03^{\rmn h} 48^{\rmn m} 43\fs143$ & $-00\degr 27\arcmin 31\farcs36$ & 20.92 & 0.3450 & 4 & \nodata & $\checkmark$ & 0.35\\
215 & 1 & $03^{\rmn h} 48^{\rmn m} 36\fs792$ & $-00\degr 27\arcmin 12\farcs79$ & 20.54 & 0.3505 & 4 & \nodata & $\checkmark$ & 0.38\\
255 & 4 & $03^{\rmn h} 48^{\rmn m} 38\fs949$ & $-00\degr 26\arcmin 50\farcs01$ & 21.29 & 0.3443 & 4 & \nodata & $\checkmark$ & 0.43\\
181 & 4 & $03^{\rmn h} 48^{\rmn m} 34\fs662$ & $-00\degr 29\arcmin 07\farcs57$ & 21.34 & 0.3570 & 4 & \nodata & \nodata & \nodata\\
165 & 3 & $03^{\rmn h} 48^{\rmn m} 33\fs621$ & $-00\degr 28\arcmin 04\farcs90$ & 20.68 & 0.3446 & 4 & \nodata & $\checkmark$ & 0.44\\
358 & 3 & $03^{\rmn h} 48^{\rmn m} 45\fs127$ & $-00\degr 27\arcmin 45\farcs73$ & 20.81 & 0.2938 & 4 & $\checkmark$ & \nodata & \nodata\\
272 & 2 & $03^{\rmn h} 48^{\rmn m} 39\fs430$ & $-00\degr 26\arcmin 39\farcs99$ & 20.69 & 0.1803 & 3 & $\checkmark$ & \nodata & \nodata\\
237 & 1 & $03^{\rmn h} 48^{\rmn m} 37\fs972$ & $-00\degr 26\arcmin 40\farcs69$ & 20.56 & 0.3459 & 4 & \nodata & $\checkmark$ & 0.49\\
153 & 2 & $03^{\rmn h} 48^{\rmn m} 32\fs984$ & $-00\degr 27\arcmin 17\farcs87$ & 22.25 & 0.4894 & 3 & \nodata & \nodata & \nodata\\
206 & 1 & $03^{\rmn h} 48^{\rmn m} 36\fs276$ & $-00\degr 30\arcmin 01\farcs29$ & 20.94 & 0.3456 & 4 & \nodata & $\checkmark$ & 0.57\\
169 & 3 & $03^{\rmn h} 48^{\rmn m} 33\fs978$ & $-00\degr 30\arcmin 01\farcs20$ & 20.46 & 0.2963 & 4 & \nodata & \nodata & \nodata\\
320 & 1 & $03^{\rmn h} 48^{\rmn m} 42\fs708$ & $-00\degr 25\arcmin 44\farcs78$ & 20.20 & 0.3602 & 3 & \nodata & \nodata & \nodata\\
371 & 3 & $03^{\rmn h} 48^{\rmn m} 45\fs867$ & $-00\degr 30\arcmin 25\farcs66$ & 21.75 & 0.3342 & 4 & \nodata & \nodata & \nodata\\
149 & 4 & $03^{\rmn h} 48^{\rmn m} 32\fs871$ & $-00\degr 30\arcmin 22\farcs95$ & 21.29 & 0.3400 & 4 & \nodata & $\checkmark$ & 0.79\\
270 & 4 & $03^{\rmn h} 48^{\rmn m} 39\fs412$ & $-00\degr 25\arcmin 30\farcs78$ & 19.13 & 0.3516 & 4 & \nodata & $\checkmark$ & 0.82\\
98 & 3 & $03^{\rmn h} 48^{\rmn m} 28\fs95$ & $-00\degr 29\arcmin 10\farcs64$ & 20.20 & 0.3412 & 4 & \nodata & $\checkmark$ & 0.83\\
107 & 3 & $03^{\rmn h} 48^{\rmn m} 29\fs523$ & $-00\degr 29\arcmin 31\farcs57$ & 20.86 & 0.3084 & 4 & \nodata & \nodata & \nodata\\
231 & 1 & $03^{\rmn h} 48^{\rmn m} 37\fs854$ & $-00\degr 25\arcmin 21\farcs24$ & 20.67 & 0.3515 & 4 & \nodata & $\checkmark$ & 0.88\\
258 & 1 & $03^{\rmn h} 48^{\rmn m} 39\fs012$ & $-00\degr 31\arcmin 35\farcs60$ & 20.27 & 0.3891 & 4 & \nodata & \nodata & \nodata\\
401 & 1 & $03^{\rmn h} 48^{\rmn m} 48\fs183$ & $-00\degr 31\arcmin 03\farcs08$ & 20.82 & 0.4575 & 4 & \nodata & \nodata & \nodata\\
141 & 1 & $03^{\rmn h} 48^{\rmn m} 32\fs555$ & $-00\degr 31\arcmin 24\farcs45$ & 19.99 & 0.2950 & 4 & \nodata & \nodata & \nodata\\
136 & 1 & $03^{\rmn h} 48^{\rmn m} 32\fs246$ & $-00\degr 25\arcmin 03\farcs39$ & 20.76 & 0.3412 & 3 & \nodata & \nodata & \nodata\\
174 & 2 & $03^{\rmn h} 48^{\rmn m} 34\fs196$ & $-00\degr 31\arcmin 48\farcs88$ & 21.63 & 0.4168 & 3 & $\checkmark$ & \nodata & \nodata\\
99 & 3 & $03^{\rmn h} 48^{\rmn m} 29\fs063$ & $-00\degr 25\arcmin 15\farcs83$ & 20.34 & 0.3422 & 3 & \nodata & \nodata & \nodata\\
\hline
\end{tabular}
\end{table*}

\begin{figure*}
\includegraphics[width=17.5cm]{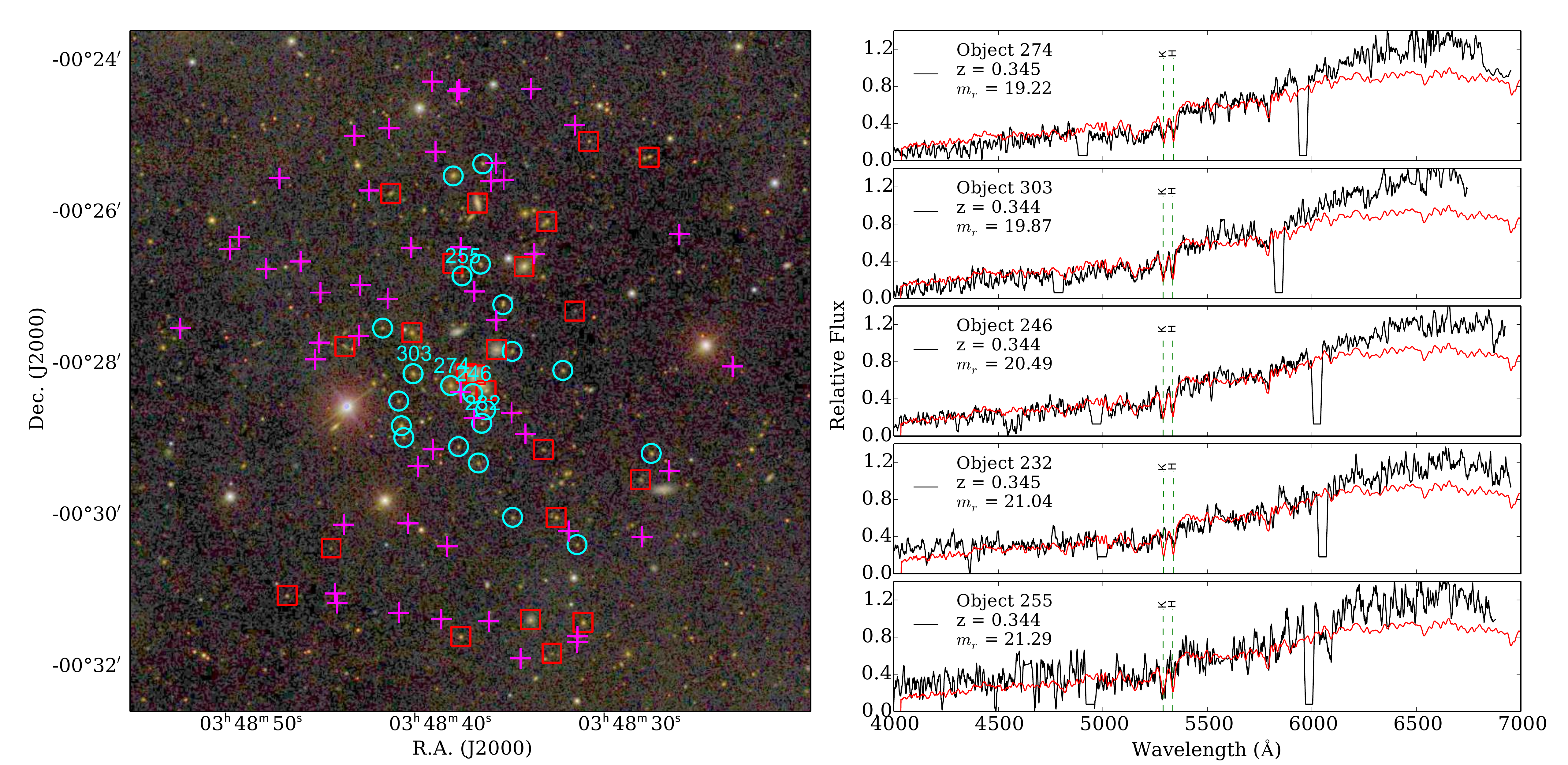}
\caption{The $z = 0.35$ cluster ACT-CL J0348.6-0028 (see Fig.~\ref{f_J0320ImageAndSpectra} for an explanation
of symbols and colours).}
\label{f_J0348ImagesAndSpectra}
\end{figure*}

\begin{table*}
\caption{Spectroscopic redshifts of galaxies in the direction of ACT-CL J2058.8+0123 measured using SALT RSS;
see Table~\ref{t_members_J0320.4+0032} for an explanation of the table columns.}
\label{t_members_J2058.8+0123}
\begin{tabular}{|c|c|c|c|c|c|c|c|c|c|}
\hline
ID & Mask & RA (J2000) & Dec. (J2000) & $m_{r}$ & $z$ & $Q$ & Em. Lines? & Member? & $r$ (Mpc)\\
\hline
219 & 2 & $20^{\rmn h} 58^{\rmn m} 56\fs777$ & $+01\degr 22\arcmin 47\farcs58$ & 19.66 & 0.3383 & 4 & \nodata & \nodata & \nodata\\
225 & 3 & $20^{\rmn h} 58^{\rmn m} 57\fs187$ & $+01\degr 21\arcmin 51\farcs00$ & 19.72 & 0.3207 & 4 & \nodata & $\checkmark$ & 0.16\\
194 & 1 & $20^{\rmn h} 58^{\rmn m} 54\fs089$ & $+01\degr 22\arcmin 24\farcs07$ & 20.71 & 0.3227 & 3 & \nodata & $\checkmark$ & 0.27\\
173 & 2 & $20^{\rmn h} 58^{\rmn m} 52\fs683$ & $+01\degr 22\arcmin 14\farcs21$ & 19.86 & 0.2043 & 4 & \nodata & \nodata & \nodata\\
211 & 1 & $20^{\rmn h} 58^{\rmn m} 55\fs861$ & $+01\degr 21\arcmin 03\farcs94$ & 20.73 & 0.3270 & 3 & \nodata & $\checkmark$ & 0.39\\
201 & 2 & $20^{\rmn h} 58^{\rmn m} 54\fs572$ & $+01\degr 20\arcmin 59\farcs56$ & 20.54 & 0.3293 & 4 & \nodata & $\checkmark$ & 0.45\\
184 & 1 & $20^{\rmn h} 58^{\rmn m} 53\fs730$ & $+01\degr 23\arcmin 36\farcs04$ & 20.83 & 0.3148 & 4 & \nodata & \nodata & \nodata\\
164 & 2 & $20^{\rmn h} 58^{\rmn m} 52\fs060$ & $+01\degr 21\arcmin 40\farcs62$ & 19.50 & 0.3334 & 4 & \nodata & $\checkmark$ & 0.46\\
213 & 1 & $20^{\rmn h} 58^{\rmn m} 56\fs096$ & $+01\degr 20\arcmin 41\farcs27$ & 20.44 & 0.3286 & 4 & \nodata & \nodata & \nodata\\
203 & 2 & $20^{\rmn h} 58^{\rmn m} 54\fs836$ & $+01\degr 20\arcmin 35\farcs18$ & 18.00 & 0.3311 & 4 & $\checkmark$ & \nodata & \nodata\\
177 & 1 & $20^{\rmn h} 58^{\rmn m} 53\fs056$ & $+01\degr 24\arcmin 10\farcs76$ & 18.13 & 0.3301 & 4 & \nodata & $\checkmark$ & 0.61\\
157 & 2 & $20^{\rmn h} 58^{\rmn m} 51\fs547$ & $+01\degr 23\arcmin 54\farcs82$ & 21.00 & 0.3260 & 4 & \nodata & $\checkmark$ & 0.62\\
200 & 1 & $20^{\rmn h} 58^{\rmn m} 54\fs514$ & $+01\degr 24\arcmin 26\farcs77$ & 20.55 & 0.3317 & 4 & \nodata & $\checkmark$ & 0.63\\
137 & 3 & $20^{\rmn h} 58^{\rmn m} 50\fs390$ & $+01\degr 23\arcmin 56\farcs26$ & 19.60 & 0.3265 & 4 & \nodata & $\checkmark$ & 0.69\\
166 & 2 & $20^{\rmn h} 58^{\rmn m} 52\fs163$ & $+01\degr 24\arcmin 30\farcs79$ & 20.38 & 0.3239 & 4 & \nodata & $\checkmark$ & 0.72\\
162 & 2 & $20^{\rmn h} 58^{\rmn m} 51\fs745$ & $+01\degr 24\arcmin 45\farcs88$ & 20.97 & 0.3281 & 4 & \nodata & $\checkmark$ & 0.80\\
190 & 1 & $20^{\rmn h} 58^{\rmn m} 53\fs906$ & $+01\degr 25\arcmin 24\farcs84$ & 19.60 & 0.1856 & 4 & $\checkmark$ & \nodata & \nodata\\
146 & 2 & $20^{\rmn h} 58^{\rmn m} 50\fs917$ & $+01\degr 25\arcmin 22\farcs66$ & 19.69 & 0.3228 & 4 & $\checkmark$ & $\checkmark$ & 0.98\\
150 & 1 & $20^{\rmn h} 58^{\rmn m} 51\fs379$ & $+01\degr 19\arcmin 14\farcs61$ & 19.92 & 0.3222 & 4 & $\checkmark$ & \nodata & \nodata\\
90 & 3 & $20^{\rmn h} 58^{\rmn m} 46\fs076$ & $+01\degr 24\arcmin 52\farcs59$ & 18.14 & 0.2935 & 4 & $\checkmark$ & \nodata & \nodata\\
153 & 1 & $20^{\rmn h} 58^{\rmn m} 51\fs408$ & $+01\degr 25\arcmin 57\farcs05$ & 19.66 & 0.3321 & 4 & \nodata & $\checkmark$ & 1.10\\
210 & 2 & $20^{\rmn h} 58^{\rmn m} 55\fs832$ & $+01\degr 26\arcmin 29\farcs51$ & 20.27 & 0.1346 & 4 & \nodata & \nodata & \nodata\\
136 & 1 & $20^{\rmn h} 58^{\rmn m} 50\fs317$ & $+01\degr 26\arcmin 09\farcs59$ & 21.34 & 0.3249 & 4 & $\checkmark$ & $\checkmark$ & 1.19\\
165 & 3 & $20^{\rmn h} 58^{\rmn m} 52\fs163$ & $+01\degr 26\arcmin 23\farcs63$ & 17.86 & 0.1344 & 4 & \nodata & \nodata & \nodata\\
\hline
\end{tabular}
\end{table*}

\begin{figure*}
\includegraphics[width=17.5cm]{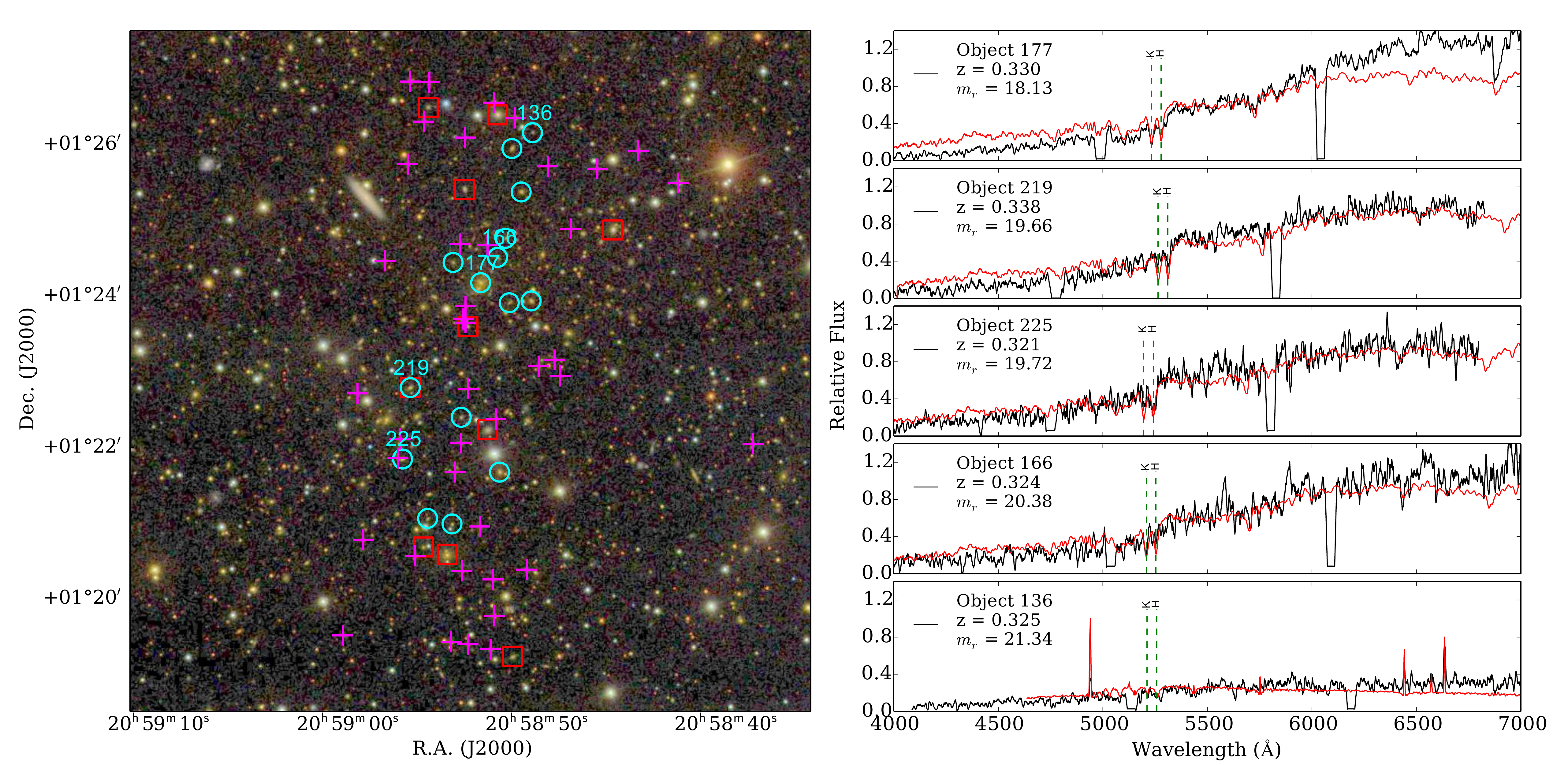}
\caption{The $z = 0.33$ cluster ACT-CL J2058.8+0123 (see Fig.~\ref{f_J0320ImageAndSpectra} for an explanation
of symbols and colours).}
\label{f_J2058ImagesAndSpectra}
\end{figure*}

\label{lastpage}

\end{document}